\documentclass[letterpaper,twocolumn,10pt]{article}
\usepackage{usenix,epsfig,endnotes}

\usepackage{enumitem}
\usepackage{algorithm}

\usepackage{algpseudocode}
\algtext*{EndWhile} %
\algtext*{EndIf} %
\algtext*{EndFor} %

\usepackage{subcaption}
\usepackage{tabularx}
\usepackage{makecell}
\usepackage{booktabs} 
\usepackage{multirow}
\usepackage{amsmath} 
\usepackage{bm}

\usepackage{titlesec}
\titlespacing*{\paragraph}{0pt}{0.5ex plus 0.2ex minus 0.1ex}{0.8em}

\usepackage{tcolorbox}
\newtcolorbox{takeaway}{
  colback=gray!10,
  colframe=gray!50,
  boxrule=0.3pt,
  arc=2pt,
  left=6pt,
  right=6pt,
  top=4pt,
  bottom=4pt,
  before skip=2pt,
  after skip=2pt
}

\usepackage{enumitem}
\setlist[itemize]{topsep=2pt, itemsep=2pt, parsep=0pt, partopsep=0pt}

\usepackage{titlesec}

\titlespacing*{\section}{0pt}{*1}{*0.7}
\titlespacing*{\subsection}{0pt}{*0.8}{*0.5}
\titlespacing*{\subsubsection}{0pt}{*0.6}{*0.4}

\begin{document}

\date{}

\title{\Large \bf Enabling Disaggregated Multi-Stage MLLM Inference via GPU-Internal Scheduling and Resource Sharing}

\author{
{\rm Lingxiao Zhao$^{\dag}$ \quad\quad Haoran Zhou$^{\dag}$ \quad\quad Yuezhi Che \quad\quad Dazhao Cheng \vspace{0.3em}} \\ 
{Wuhan University}
}

\maketitle
\def\thefootnote{\dag}\footnotetext{Equal contribution}\def\thefootnote{\arabic{footnote}}

\begin{center}
{\large \bf Abstract}
\end{center}

Multimodal large language models (MLLMs) extend LLMs with visual understanding and rely on a three-stage pipeline consisting of multimodal preprocessing, vision encoding, and LLM inference.~While integrating these additional stages enriches model capability, we identify critical system bottlenecks. 
First, heavy multimodal preprocessing, particularly video decoding, frequently dominates the Time-to-First-Token (TTFT). 
Most deployments default to CPU-based decoding, but this severely limits throughput. While existing GPU decoding enables throughput-oriented parallelism, it does not align with the latency-sensitive needs of the MLLM pipeline.
Second, after decoding, the subsequent vision encoder acts as an independent and compute-intensive model that transforms visual inputs into embeddings. Such heterogeneous computation cannot be co-batched with LLM prefill or decode, forcing inter-stage blocking that increases token-generation latency. Even when these stages are executed on separate GPUs to avoid blocking, the system cannot fully utilize available compute and memory resources because each stage operates on only a subset of the hardware, lowering overall utilization and constraining throughput.

To address these challenges, we present FlashCodec and UnifiedServe, two complementary designs that jointly optimize the end-to-end MLLM pipeline. 
FlashCodec accelerates the multimodal preprocessing stage through collaborative multi-GPU video decoding, reducing decoding latency while preserving high throughput.
UnifiedServe optimizes the vision-to-text and inference stages using a logically decoupled their execution to eliminate inter-stage blocking, yet physically sharing GPU resources to maximize GPU system utilization. By carefully orchestrating execution across stages and minimizing interference, UnifiedServe Together, our proposed framework forms an end-to-end optimized stack that can serve up to 3.0$\times$ more requests or enforce 1.5$\times$ tighter SLOs, while achieving up to 4.4$\times$ higher throughput compared to state-of-the-art systems.

\section{Introduction}
MLLMs, such as Gemini-1.5~\cite{team2024gemini}, GPT-4o~\cite{hurst2024gpt}, and open-source series like
LLaVA~\cite{li2024llava_next, li2024llava_one, huang2024dynamic}, Qwen-VL~\cite{bai2025qwen2, wang2024qwen2, Qwen3-VL}, and InternVL~\cite{wang2025internvl3, zhu2025internvl3} extend the intelligence of LLMs from text to images and videos, enriching AI applications toward diverse multimodal interactions.
The frontier of MLLM capability has shifted from image understanding~\cite{chen2022visualgpt, hu2023promptcap, schwenk2022okvqa, shao2023prompting} to comprehensive video reasoning and analytics such as action recognition~\cite{caba2015activitynet, goyal2017something, kay2017kinetics}, visual object tracking~\cite{huang2019got, muller2018trackingnet}, and video question-answering~\cite{xu2017video, jang2017tgif, xiao2021next, li2024mvbench}.
As MLLMs grow in capability and are rapidly deployed in real production systems~\cite{qiu2025modserve}, the need to meet latency and throughput requirements makes efficient MLLM serving increasingly critical.
Serving MLLMs follows a Vision-Text-to-Text~\cite{Image-Text-to-Text, qiu2025modserve} inference pipeline composed of three stages. First, raw images or video frames are decoded into tensor representations, referred to as patch tokens\footnote{Patch tokens denote the tensor representations produced by the multimodal preprocessing stage.}. Second, a vision encoder processes these patch tokens into visual embeddings compatible with the LLM. Third, the LLM integrates visual embeddings with textual inputs to generate responses. The first two vision-related stages serve as the bridge that empowers text-only LLMs with visual modality.
However, introducing these heterogeneous stages fundamentally alters the serving workflow and introduces critical performance bottlenecks:

First, heavy video decoding frequently dominates the TTFT, as recent studies show that multimodal preprocessing can account for the majority of early-stage latency under video workloads~\cite{shao2025tokens,liu2025elasticmm}.
Most mainstream deployments rely on CPU-based decoding~\cite{dong2025hydrainfer, singh2024efficiently, kwon2023efficient, zheng2024sglang}, as common codecs such as H.264 decode efficiently on CPUs and naturally avoid interference with GPU execution, yet CPU throughput remains low and scales poorly for large videos.
Meanwhile, GPU decoders are designed for high-throughput multi-stream processing and offer limited improvement for the latency-critical path required by MLLM serving.

Second, the vision encoding stage introduces a compute-intensive and heterogeneous workload that fundamentally disrupts the flow of LLM inference.
To understand how current systems cope with this additional stage, we investigate existing MLLM serving designs and find two main paradigms:~(\textit{i})~\textit{Monolithic}-based and (\textit{ii})~\textit{Split}-based scheduling designs.
\textit{Monolithic} designs co-locate the encoder and the LLM within a single service instance~\cite{aminabadi2022deepspeed, kwon2023efficient, zheng2024sglang}, allowing all stages to utilize the full set of GPU resources~(e.g., compute and memory) and thus maximizing overall system throughput. Yet its Time Between Tokens~(TBT)~Service Level Objective~(SLO) is often violated due to interference introduced by the encoder stage.
In contrast, \textit{Split} services adopt a Prefill-Decode~(PD)-disaggregation architecture~\cite{qiu2025modserve, dong2025hydrainfer, singh2024efficiently}, deploying the encoder as an independent instance, avoiding cross-stage interference but fragmenting compute and memory resources across the GPUs. 
As a result, \textit{Split} design reduces the overall system throughput, lowering the aggregate tokens-per-second across all users.

\begin{figure}
    \centering
    \begin{subfigure}{0.49\linewidth}
        \centering
        \includegraphics[width=\linewidth]{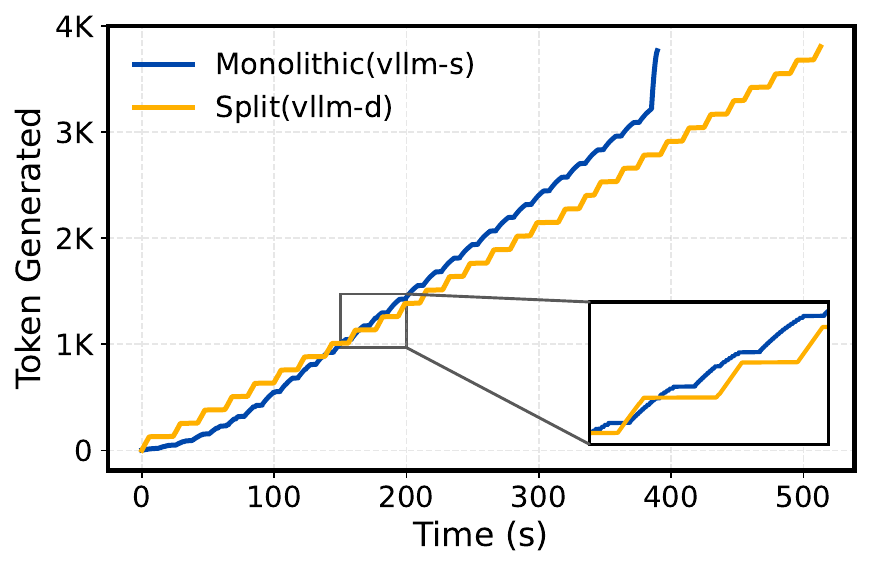}
         \vspace{-5mm}
        \caption{Token-generation progress.}
        \label{intro1}
         \vspace{-2mm}
    \end{subfigure}
    \begin{subfigure}{0.49\linewidth}
        \centering
        \includegraphics[width=\linewidth]{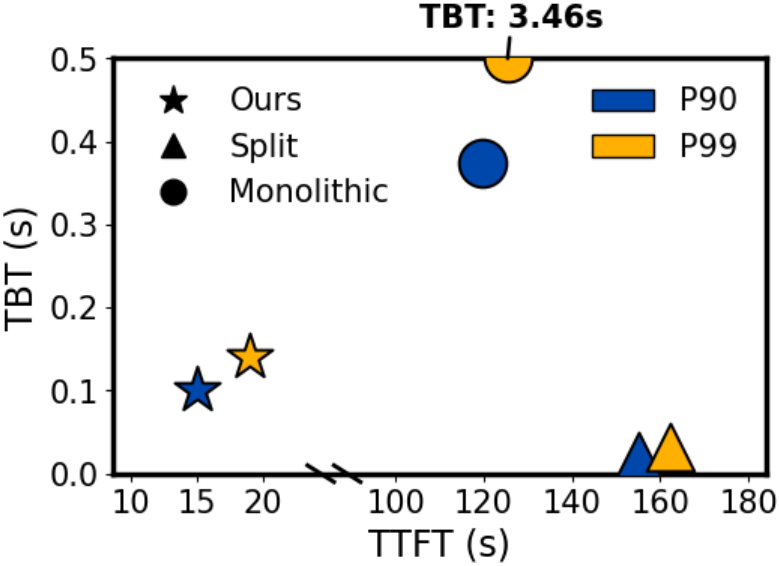}
         \vspace{-5mm}
        \caption{P90/P99 TTFT and TBT.}
        \label{intro2}
        \vspace{-2mm}
    \end{subfigure}
    \caption{
    Qwen2.5-VL-32B running on four A100 GPUs serving 30 video requests from MLVU dataset. (a) Both systems experienced frequent generation stalls. (b) The overall performance across all systems
    }
    \vspace{-5mm}
    \label{intro}
\end{figure}

Figure~\ref{intro1} illustrates the token-generation progress over time under \textit{Monolithic} and \textit{Split} serving architectures. 
Both architectures exhibit visible stalls (plateaus), yet for fundamentally different reasons.
For \textit{Monolithic}-based services, numerous stalls occur during generation due to interference from the encoder. 
While the \textit{Split} design produces a smoother because the first-token path, which includes video decoding and vision encoding, takes substantially longer than the text-only LLMs and cannot keep up with LLM decode consumption, ultimately leading to generation stalls.
Figure~\ref{intro2} quantifies these effects through the P90/P99 TTFT and TBT metrics. The \textit{Split} architecture achieves the lowest and most stable TBT, whereas its TTFT is noticeably slower. Meanwhile, the \textit{Monolithic} architecture shows significantly inflated P99 tail latency, reflecting the severe generation stalls observed in Figure~\ref{intro1}. These complementary trade-offs clearly empirically validate our analysis. 

\textbf{\textit{How to achieve the low TTFT and TBT simultaneously?}} 
As shown in Figure~\ref{intro2}, our proposed design breaks this dilemma, achieving the sweet spot of low latency on both metrics in this case. This outcome stems from two insights:

First, in Vision-Text-to-Text models, heavy video inputs make vision preprocessing a dominant contributor to initial latency. CPUs are limited by scaling, while current GPU decoders process distinct videos in parallel but ignore single-request latency.
By collaboratively decoding one video across all GPUs and fully exploiting all available hardware decoding engines, the system can significantly reduce TTFT.

Second, we can logically decouple the MLLM stages while physically enabling resource sharing across GPUs. Instead of rigid PD-disaggregation or monolithic blocking, our architecture treats the entire GPU cluster as a shared resource pool. This allows each stage to execute independently, greatly reducing cross-stage interference. Therefore, the system achieves higher utilization and sustains low-latency execution even under heavy multimodal workloads.

Accordingly, we develop \textbf{FlashCodec} and \textbf{UnifiedServe}.
FlashCodec accelerates multimodal preprocessing by exploiting all GPUs in the system and the full multimedia-decoding capability~(e.g., NVDEC, JPEG-decode) available on each device. 
For video inputs, FlashCodec enables fine-grained parallelism by partitioning a video into independently decodable, non-redundant segments and dispatching them across GPUs and NVDEC engines within each GPU. It further employs a stall-free scheduling strategy that eliminates NVDEC idle gaps and ensures continuous decoding progress.
UnifiedServe orchestrates the inference as three asynchronous yet coordinated workers: a vision-preprocess worker, an encode-prefill worker, and an LLM decode worker. 
The vision-preprocess worker employs FlashCodec to decode multimodal inputs. 
The encode-prefill worker performs encoding and prefill in a mutually blocking manner to bound resource contention, while the LLM decode worker runs in a separate process to ensure low-latency TBT. 
To handle the data dependencies between these asynchronous stages, UnifiedServe incorporates an efficient buffering mechanism that manages multimodal intermediate states without excessive memory overhead, enabling high system utilization while preserving low latency.

In summary, we make the following contributions:
\begin{itemize}[left=0pt]
    \item We conduct an in-depth analysis of performance bottlenecks in MLLM serving. We quantitatively validate these bottlenecks and characterize the trade-offs between existing \textit{Monolithic}- and \textit{Split}-based serving architectures.
    
    \item We propose FlashCodec, a collaborative multi-GPU video decoding mechanism for high-throughput, low-latency multimodal preprocessing.
    
    \item We introduce UnifiedServe, a serving architecture that logically decouples the MLLM pipeline while physically enabling full-system resource sharing.
    
    \item Our proposed framework forms an end-to-end optimized stack that can serve up to 3.0$\times$ and achieving up to 4.4$\times$ higher throughput compared to state-of-the-art systems.
\end{itemize}

\section{Background and Motivation}

\begin{figure}
    \centering
    \includegraphics[width=1\linewidth]{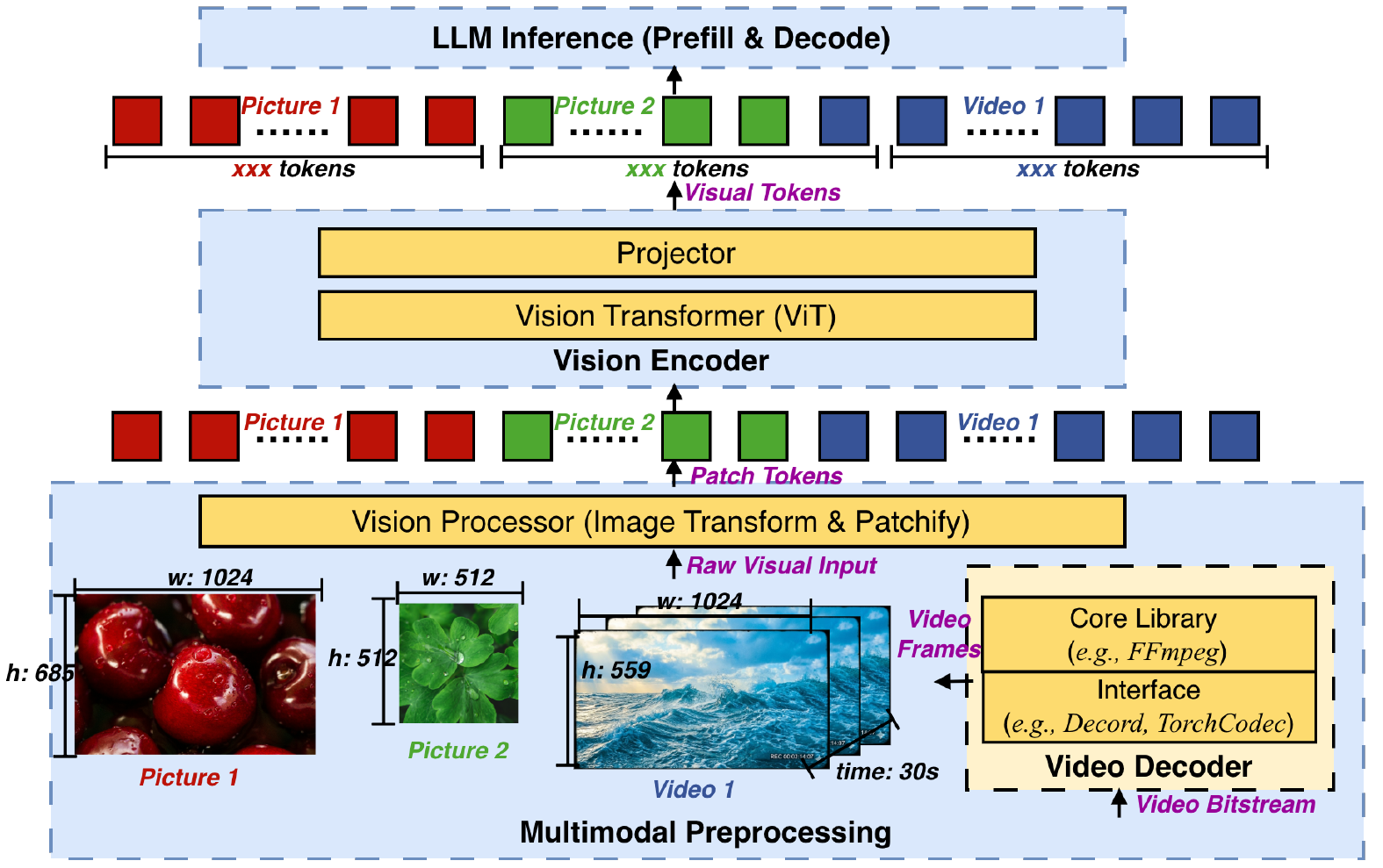}
    \vspace{-2mm}
    \caption{MLLM inference process overview.}
    \label{fig:bg-over}
    \vspace{-4mm}
\end{figure}

\subsection{MLLM Inference Overview}
A typical MLLM inference workflow consists of three heterogeneous stages, as shown in Figure~\ref{fig:bg-over}: 
\emph{(i) Multimodal input preprocessing.} The process begins by decoding compressed images/videos into raw pixel-space frames, which are then transformed into uniformly sized tiles (patch tokens) that serve as the input to the next phase. 
\emph{(ii) Vision Encoder.}~These patch tokens are then  passed to a vision encoder, which converts them into high-level embeddings (visual tokens) suitable for LLM input.
\emph{(iii) LLM inference.}~The LLM backend concatenates the visual and text tokens, performs a \textit{prefill} computation to construct the KV cache, and then enters an autoregressive \textit{decode} phase that generates output tokens sequentially.
These heterogeneous stages form a substantially more complex end-to-end execution path.

\subsection{Toward Low-Latency, High-Throughput Video Decoding for MLLM Serving}\label{mtv_vision_decoding}

Meeting SLOs for MLLM serving requires resolving the latency bottleneck introduced by video decoding.
Before entering the vision encoder, image and video inputs must be decoded into raw pixel formats (e.g., RGB). 
Unlike lightweight image decoding, video decoding is substantially more complex and time-consuming, frequently emerging as the dominant contributor to end-to-end inference latency~\cite{schneider2025quickvideo}.
Contemporary ML video-decoding frameworks (e.g., OpenCV~\cite{opencv_decoder}, Decord~\cite{decord_decoder} and TorchCodec~\cite{torchcodec_decoder})~typically invoke FFmpeg~\cite{ffmpeg} in the backend responsible for the actual decoding. Technical details of video decoding are introduced in Section~\ref{video_de_bg}.

\begin{figure}
    \centering
    \includegraphics[width=0.85\linewidth]{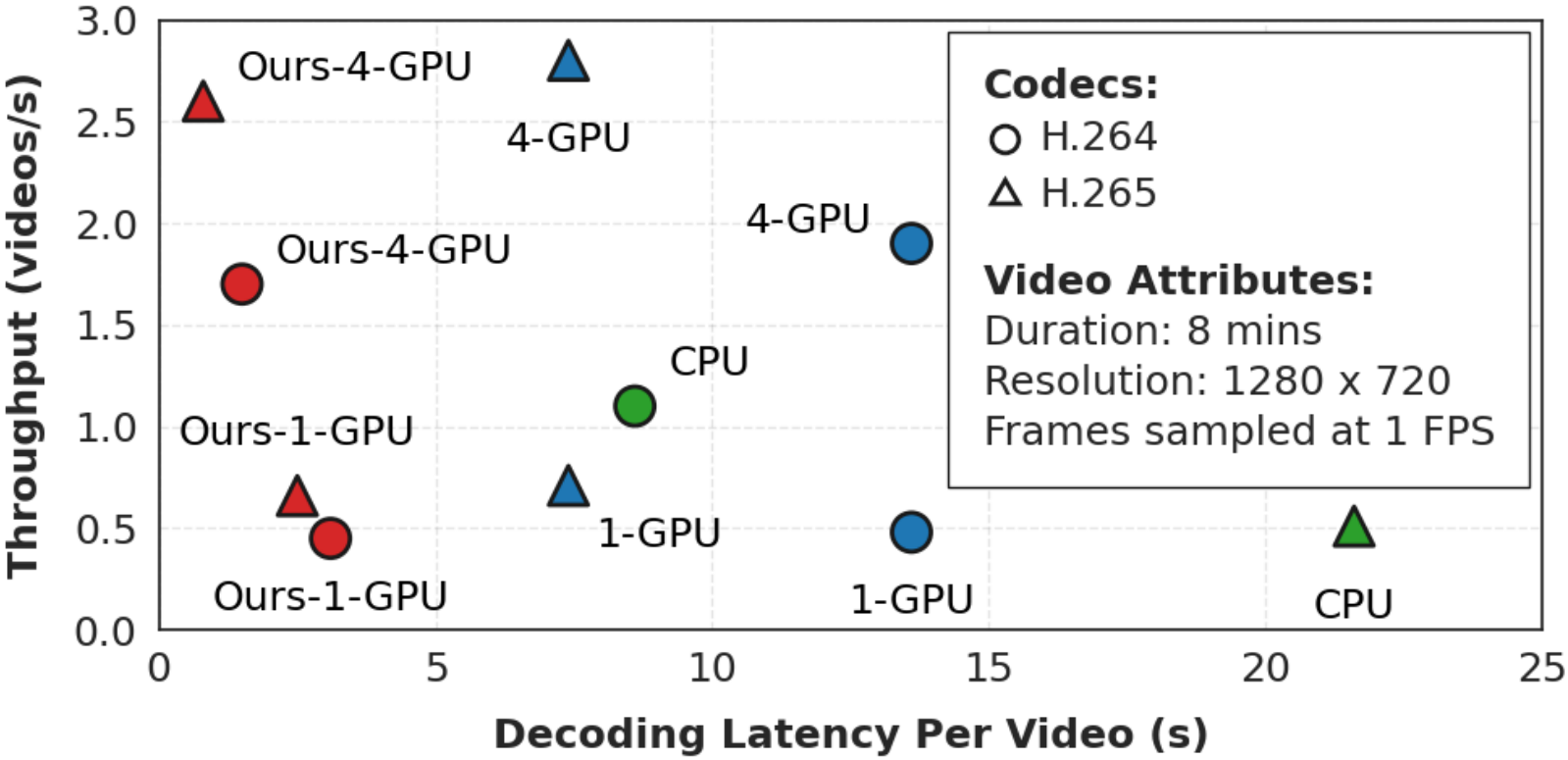}
    \vspace{-2mm}
    \caption{Performance comparison across different configurations, with Decord~\cite{decord_decoder} as the default CPU/GPU decoder.}
    \label{fig:mtv_video_decoding_2_2_1}
    \vspace{-3mm}
\end{figure}

\paragraph{Limitations of Current CPU and GPU Decoding.} 
Although these frameworks support both CPU decoding and GPU hardware decoders such as NVDEC, mainstream MLLM inference frameworks (e.g., SGLang~\cite{zheng2024sglang} and vLLM~\cite{kwon2023efficient}), and even production MLLM inference deployments~\cite{qiu2025modserve}, predominantly default to CPU-based decoding. 
Our analysis suggests that this design choice is driven by codec characteristics and GPU resource contention. 
In practice, H.264~\cite{h264} remains the dominant codec in real-world workloads, and remains comparatively CPU-friendly. 
As shown in Figure~\ref{fig:mtv_video_decoding_2_2_1}, server-grade CPUs can decode non–high-resolution H.264 videos (e.g., $< 1$K) faster than default GPU decoding in single-request scenarios.
Moreover, CPU-based decoding naturally avoids interference with GPU compute and memory activity, preventing contention with model inference and simplifying GPU memory management.

However, despite these advantages, the absolute latency of CPU decoding remains high and degrades further for non-H.264 codecs (e.g., H.265) or videos with higher resolution or longer duration.
Also, GPU decoding already outperforms the CPU for H.265 videos, as shown in Figure~\ref{fig:mtv_video_decoding_2_2_1}; and GPUs offer high throughput by decoding multiple videos in parallel, making them appealing for throughput-oriented workloads such as pretraining vision foundation models~\cite{DALI}. Yet this parallelism does not reduce per-video decoding latency, and neither CPU decoding nor GPU decoding can meet the strict TTFT requirements of SLOs.

\paragraph{Collaborative GPU Decoding is the Key.}
Instead of decoding distinct videos in parallel, we find that allocating multiple GPU decoding resources to collaboratively decode a single video can significantly reduce decoding latency and improve the end-to-end SLO performance. 
Figure~\ref{fig:mtv_video_decoding_2_2_1} shows that our collaborative decoding achieves $2.8$-$4.4\times$ speedup when utilizing all NVDEC engines on one A100 GPU for a $1280\times720$ H.264 and H.265 video, outperforming both CPU and default GPU paths.
Scaling the decoding across four GPUs further improves performance to $5.7$-$9.1\times$, bringing single-video latency into the sub-second regime while maintaining relatively high throughput. 

\begin{takeaway}
\noindent\textbf{Takeaway-1:}
\textit{
Collaborative use of all GPU decoding resources 
can substantially reduce per-video decoding latency while preserving high throughput.
}
\end{takeaway}

\begin{figure}
    \centering
    \includegraphics[width=1\linewidth]{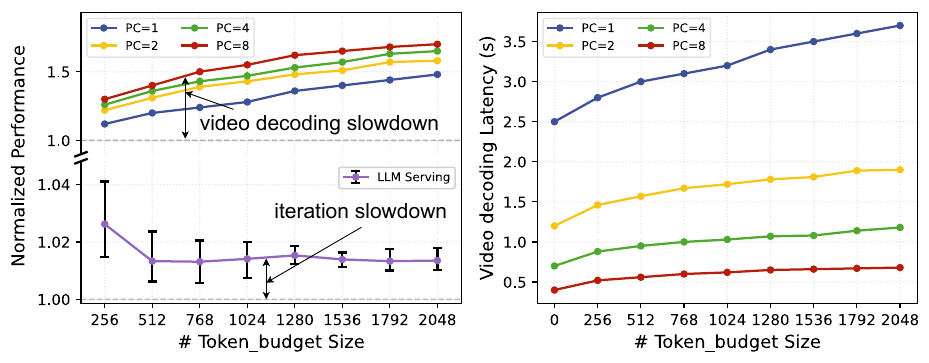}
    \caption{
    \textbf{(a)}~Normalized performance of video decoding and LLM serving under different parallel configurations (PC).
    \textbf{(b)}~Video decoding latency value. 
    }
    \label{fig:mtv_video_decoding_2_2_2}
    \vspace{-4mm}
\end{figure}

\paragraph{Overcoming Asymmetric Interference through Multi-GPU Scaling.}
To avoid blocking LLM serving, video decoding must run concurrently with inference. When decoding is performed on GPUs, resource contention between decoding and LLM computation becomes inevitable. 
Our experiments show that this interference is highly asymmetric: decoding is significantly slowed by inference activity, whereas inference is barely affected by decoding. 
Nevertheless, this slowdown can be effectively mitigated through multi-GPU scaling.
As shown in Figure~\ref{fig:mtv_video_decoding_2_2_2}~(a), increasing token budgets causes decoding latency by 48-70\%, but the per-iteration inference latency rises by under $2\%$.
Because inference typically spans multiple GPUs, decoding can also be distributed across GPUs. This scaling absorbs interference and maintains low end-to-end latency. Figure~\ref{fig:mtv_video_decoding_2_2_2}~(b) shows that decoding latency remains below $2$s with 2-4 GPUs and under $1$s with 8 GPUs.

\begin{takeaway}
\noindent\textbf{Takeaway-2:}
\textit{
Inference is resilient to decoding, while decoding is sensitive to inference. Collaborative decoding ensures low latency through multi-GPU scaling.
}
\end{takeaway}

\subsection{Toward Unified and Interference-Aware MLLM Scheduling}\label{Interference_Analysis}
After video decoding (discussed in \S\ref{mtv_vision_decoding}), the MLLM pipeline proceeds to a vision encoding stage that transforms patch tokens into visual embeddings. 
Unlike text-only LLMs, this additional compute-intensive stage must be integrated into the downstream LLM scheduling, making scheduling policy design far more challenging.
\paragraph{Scheduling Policies for LLM Inference.}
Autoregressive LLM inference exhibits a well-understood asymmetry between its two execution phases: the prefill phase is compute-bound while decoding is memory-bound. In this work, we group existing strategies into two categories: \textit{Monolithic}-based and \textit{Split}-based scheduling.

\begin{itemize}[left=0pt]
    \item \textit{Monolithic}:~Chunked-prefill~\cite{agrawal2024taming}, widely adopted in recent systems~\cite{agrawal2024taming, Deepspeed_mii}, improves upon earlier \textit{prefill-prioritizing}~\cite{yu2022orca} and \textit{decode-prioritizing}~\cite{fastertransformer, triton_dynamic_batcher} approaches. By operating at fine-grained chunk granularity, it interleaves prefill and decode within the same batch as homogeneous compute units, achieving high GPU throughput without explicitly distinguishing between stages.
    \item \textit{Split}-based: Disaggregation~\cite{zhong2024distserve, patel2024splitwise} approaches decouple prefill and decode onto separate GPUs or nodes to eliminate mutual interference, thereby stabilizing TBT to meet strict SLOs.
\end{itemize}

\begin{figure}
    \centering
    \includegraphics[width=1\linewidth]{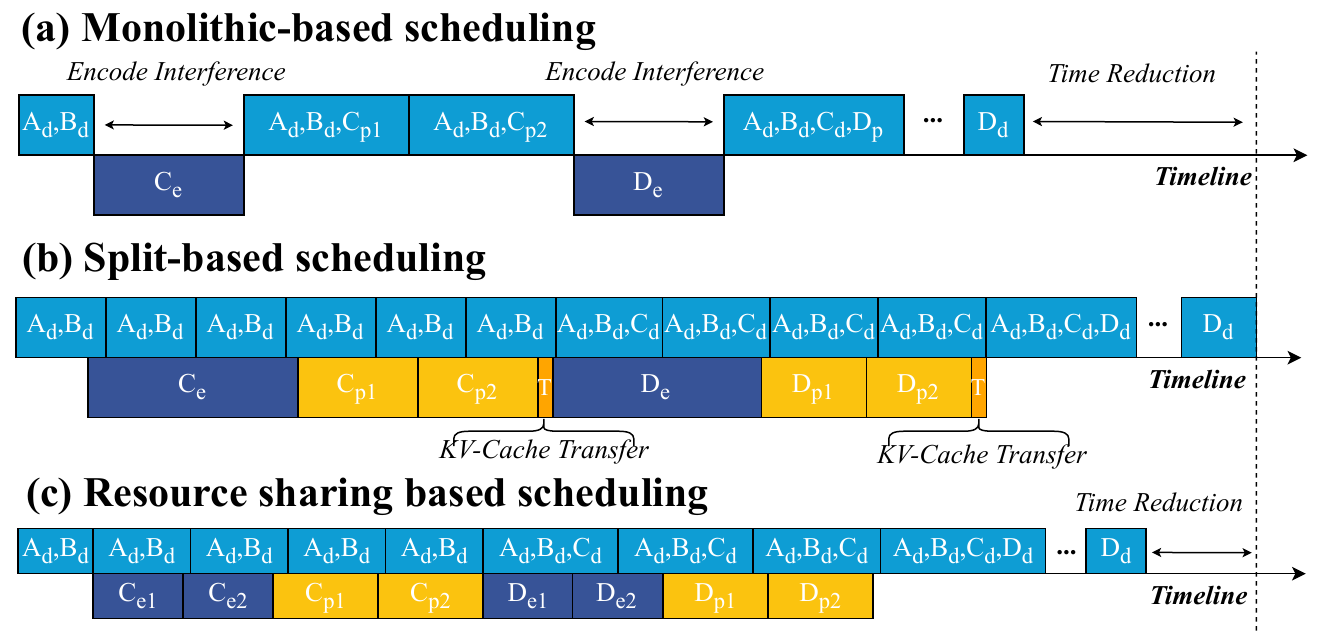}
    \caption{Comparison of three scheduling pipelines.}
    \label{fig:mtv_schedule_2_3}
    \vspace{-3mm}
\end{figure}

\paragraph{Inefficiency of Current Scheduling on MLLMs.}
Current scheduling strategies excel at text-only workloads and implicitly assume a single-model execution pipeline. 
However, vision encoding is executed by a separate model (e.g., ViT) and forms a heterogeneous and compute-intensive path~\cite{qiu2025modserve}. Unlike text prefill, which shares homogeneous operators with decoding, vision computation is heterogeneous and cannot be coalesced into ongoing decode batches for joint execution. 
As abstracted in Figure~\ref{fig:mtv_schedule_2_3}(a), when the encoder is inserted into a unified chunked schedule, we find that it monopolizes GPU compute and blocks ongoing decode tokens, which forces the scheduler into \emph{encode-prioritizing}, causing substantial TBT inflation. 
On the other hand, while effective for TBT stability, this rigid separation in \emph{Split} magnifies the inherent asymmetry between compute-bound encode/prefill and memory-bound decode. Only a subset of devices handles the heavy encoding workload, inflating TTFT under realistic load. Moreover, in this case, memory-bound nodes cannot lend their idle compute resources, nor can compute-bound nodes fully utilize their memory bandwidth for decoding~\cite{agrawal2024taming, Deepspeed_mii}. This imbalance leads to chronic resource fragmentation, causing low system utilization.

\begin{takeaway}
\noindent\textbf{Takeaway-3:}
\textit{The independent and compute-heavy vision encode path exposes a fundamental mismatch between MLLM workloads and existing LLM schedulers, necessitating a new design that can simultaneously satisfy strict SLOs while maximizing throughput.}
\end{takeaway}

\paragraph{Our Idea: Logically Decoupling the Pipeline Stages while Physically Sharing GPU Resources. }\label{Resource_Sharing_Analysis}
Ideally, this allows us to harvest the idle SM cycles during the decode phase to asynchronously execute compute-intensive vision encoding tasks. 
As shown in Figure~\ref{fig:mtv_schedule_2_3}(c), our design theoretically eliminates the blocking of \textit{Monolithic} scheduling while avoiding the resource fragmentation of \textit{Split} scheduling.
By prioritizing the decode stream while allowing the encoder to utilize leftover cycles, we enable non-blocking decoding while saturating compute units, maximizing overall throughput.

\begin{figure}
    \centering
    \includegraphics[width=1\linewidth]{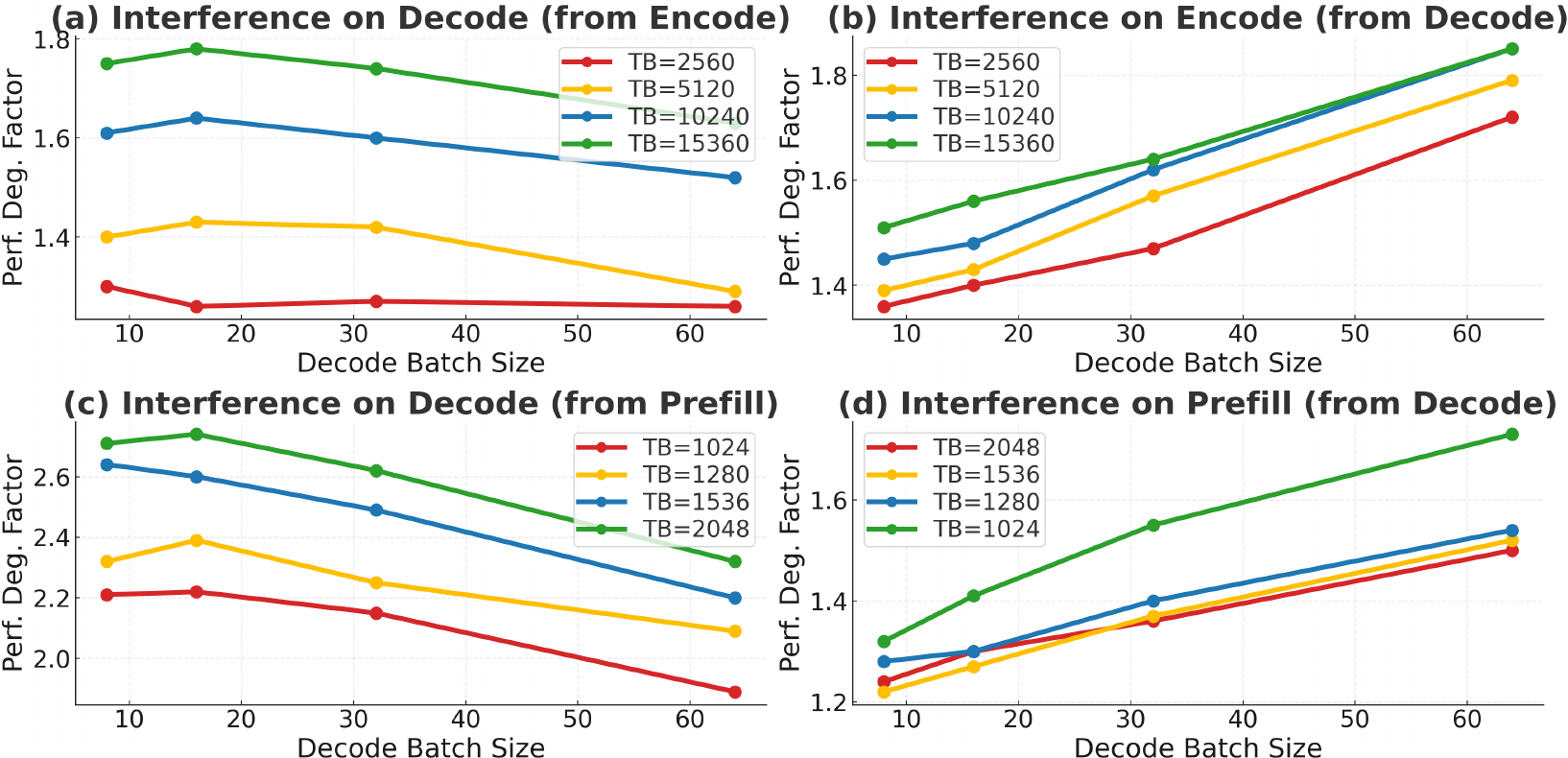}
    \caption{Mutual interference among different phases for Qwen-2.5-VL-7B~\cite{Qwen2.5-VL} running on a single 80GB A100. We use prompt length of 1024 for both prefill and decode phase. The resolution of each image is $224 \times 224$ in encode phase.}
    \label{fig:mtv_interference_2_4}
    \vspace{-3mm}
\end{figure}

\paragraph{Challenge: Characterizing Resource Contention}
While promising, resource sharing introduces resource contention. 
Figure~\ref{fig:mtv_interference_2_4} quantifies this resource-contention interference.
The \emph{Decode Batch Size} controls the compute intensity of the LM decode phase: a larger batch size increases parallelism and makes decode intensive. Similarly, the \emph{Token Budget}~(TB) of the encode or prefill phase determines their compute cost. Each value in the figure reports a \emph{Performance Degradation Factor}, defined as the slowdown of a victim phase when another phase runs concurrently.
Subfigures (a)/(c) present degradation on decode caused by encode/prefill, while (b)/(d) present degradation on encode/prefill caused by decode.
The results reveal that \emph{Compute-intensive phases dominate GPU resources}: a stage with higher compute demand not only experiences less slowdown under contention but also imposes disproportionately larger slowdown on other concurrently executing stages.
Conversely, lighter phases are more susceptible to interference, and their resources are more easily preempted by heavier ones.
In addition, encode and prefill differ in their contention strength—prefill, being more compute-heavy, exerts substantially stronger pressure on decode than encode does.

This empirical structure of contention suggests a practical control lever: tuning encode and prefill load~(e.g., via TB) indirectly regulates decode TBT under resource sharing.

\begin{takeaway}
\noindent\textbf{Takeaway-4:}
\textit{
Logically separating the multi-stage MLLM inference pipeline and executing the stages in parallel via resource sharing integrates the advantages of both split-based and unified-based scheduling.
}
\end{takeaway}

\section{FlashCodec Techniques}\label{sec::flashcodec_techniques}
FlashCodec aims to support efficient video and image decoding. 
FlashCodec leverages parallelization and stall-free scheduling to fully utilize the video and image decoding resources on the GPU while minimizing GPU memory consumption. 
Since image decoding is more straightforward than video decoding, we only discuss optimizations for video decoding.
All decoding operations are performed asynchronously (releasing the GIL during decoding), ensuring that the scheduling of decode requests is not blocked by the execution of decoding tasks.

This section reviews the key Flashcodec video decoding techniques: GOP-bsed Parallel decoding(~\S\ref{GOP}), stall-free Scheduling(~\S\ref{Stall-Free}) and memory saving method(~\S\ref{memory-saving}). Section \ref{degisn} presents the UnifiedServe system design.

\subsection{Introduction to Video Decoding.}\label{video_de_bg}
Multimedia containers such as MP4 or MKV store video as a compressed bitstream, alongside other media tracks (e.g., audio or subtitles) and the metadata needed for decoding~\cite{ebrahimi2000mpeg,wiegand2003overview}. 
During decoding, the video bitstream can be viewed as a sequence of \emph{packets}
, which are processed by the video decoder to reconstruct the ordered video frame sequence.
\emph{Packets} are not frame-aligned; decoding one packet may produce zero, one, or multiple frames. 
Moreover, frames are grouped into \emph{Group of Pictures} (\emph{GOPs}), each beginning with a self-contained keyframe (I-frame) and followed by frames (P-/B-frames) whose reconstruction depends on previously decoded content. Consequently, frames within a GOP must be decoded sequentially.

Unlike sequential video decoding for human playback, MLLMs require the specified video frames to be fully decoded upfront. For each desired frame $f_i$, identified by indices $I \subseteq \{1, 2, \ldots, m-1\}$, the decoder, upon seeking to the target index, jumps to the nearest preceding keyframe and decodes forward if the target frame lies outside the current GOP; otherwise, it decodes sequentially to the target frame~\cite{torchcodec_decoder}. Seeking outside the current GOP can incur additional overhead, as the decoder must flush internal buffers and reinitialize its reference state~\cite{ffmpeg}.

\subsection{GOP-bsed Parallel Decoding} \label{GOP}
Although inter-frame dependencies typically enforce sequential decoding in video processing, a certain degree of parallelism exists across different GOPs. This structural property enables videos to be decoded independently at the GOP level, thereby allowing parallel decoding across GOPs (\S\ref{video_de_bg}).

Algorithm~\ref{alg:GOP-based-parallel} presents the core of our GOP-based parallel video decoding. 
The algorithm first leverages the video metadata to construct the $GOP_s\_VEC$ for each GPU rank, which records, for every NVDEC unit assigned to that rank, the PTS of each frame as well as the corresponding GOP location (lines 1–3).
Video decoding starts by seeking to the start of each $GOP_s$ (line 5). 
Finally, packets are enqueued for decoding, and we select the desired frames, resize them, and convert them into GPU tensors (lines 6-26). 
In addition, for H.264 videos, seeking with the FFmpeg backend incurs substantial overhead (approximately 50-100 ms). Consequently, we perform a single initial seek to the keyframe nearest the first target frame and then decode all subsequent frames sequentially. For other codecs, we seek between GOPs and decode frames sequentially within each GOP.

\begin{figure}
    \centering
    \includegraphics[width=1\linewidth]{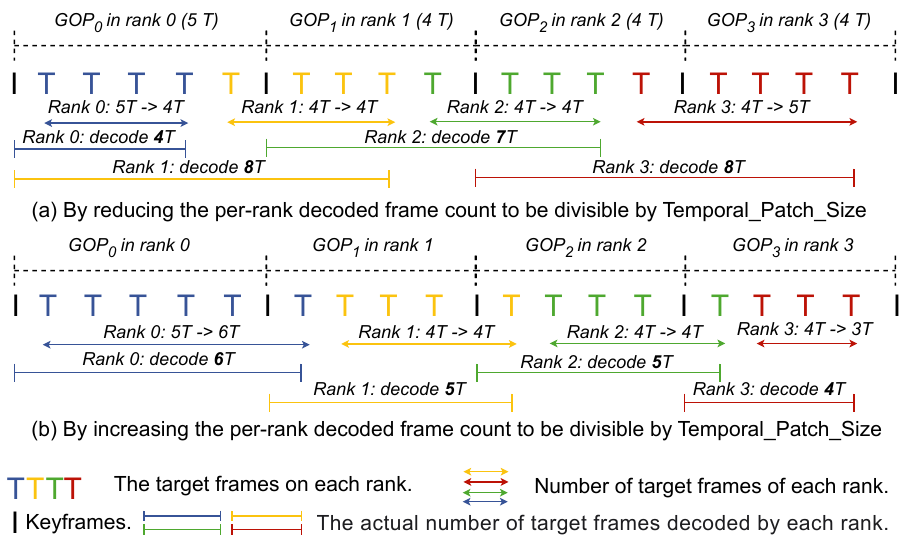}
    \caption{Two methods for making divisible by Temporal\_Patch\_Size(Assume each rank has only one GOP for brevity).Method b reduces the additional decoding overhead from $10T(27T-17T)$ to $3T(20T-17T)$.}
    \label{fig:alignment_to_tps}
    \vspace{-3mm}
\end{figure}

Some MLLMs apply temporal compression to patch tokens prior to encoding, i.e., merging $T$ (TEMPORAL\_PATCH\_SIZE) consecutive frames into a single temporal patch. If the total frame count is not divisible by $T$, the sequence is padded with the last frame to achieve divisibility \cite{Qwen2.5-VL}. Because each GPU rank processes patch tokens independently (\S\ref{sec::vsual_modality_processing}), the compression outputs must remain consistent with those produced on a single GPU.
Let the parallel world size be $W$. To ensure consistency, we adjust the frame counts of the first $W-1$ ranks, so that each is individually divisible by $T$, while padding is applied only on the final rank to make the total frame count divisible by $T$. Two adjustment strategies exist for the first $W-1$ ranks: decreasing or increasing their frame counts. Decreasing frames (Figure \ref{fig:alignment_to_tps}a) incurs nearly a full-GOP decoding penalty, since frames within a GOP must be decoded sequentially (\S\ref{video_de_bg}). Thus, we increase the per-rank frame count instead (Figure \ref{fig:alignment_to_tps}b), which minimizes additional decoding cost (line 4).

\begin{algorithm}[t]
\footnotesize
\caption{\textbf{GOP-bsed parallel decoding} with FlashCodec.In a multi-GPU setting, each rank executes this algorithm in parallel to decode different video frame portions.
}
\label{alg:GOP-based-parallel}
\begin{algorithmic}[1]
\State \textbf{Input:} Bit stream:$S$, P\_World\_Size:$W$, P\_Rank:$R$, Num\_NVDEC:$N$
\State Initialize \texttt{$M$}
$\leftarrow \texttt{get\_video\_metadata}(S)$
\State Initialize \texttt{$GOP_{s}\_VEC$}
$\leftarrow \texttt{get\_GOPs\_per\_rank}(S, W, R, N)$
\State $make\_align\_to\_temporal\_patch\_size(GOP_s, W, R)$
\State $AVFrames$
$\leftarrow alloc\_avframe\_vec(GOP_s)$
\State $Seek\_to\_keyframe\_of\_the\_first\_GOP(S, GOP_s)$
\For{$GOP_s$ in $GOP_s\_VEC$ in parallel}
\State $PTX\_index \leftarrow 0$
\For{$GOP$ in $GOP_s$}
    \If{$M.codec$ is not H.264}
        \State $Seek\_to\_nearest\_keyframe(S, GOP.first)$
    \EndIf
    \While{True}
        \State $f \leftarrow$ $decode\_one\_frame(S)$
        \If{$f.pts = GOP[PTX\_index]$}
           \State $AVFrames \leftarrow f$
           \State $PTX\_index \leftarrow PTX\_index+1$
        \EndIf
        \If{$GOP.last \leq f.pts$}
            \State $break$
        \EndIf
    \EndWhile
\EndFor
\EndFor
\State \texttt{$F$}
$\leftarrow alloc\_gpu\_tensor(M, AVFrames)$
\State \texttt{$F$}
$\leftarrow convert\_AVframes\_to\_tensor\_and\_resize(F,M, AVFrames)$
\end{algorithmic}
\end{algorithm}

\begin{algorithm}[t]
\footnotesize
\caption{\textbf{Stall-free Scheduling} with FlashCodec. In a multi-GPU setting, each rank executes this algorithm in parallel to schedule video decoding tasks.
}
\label{alg:stall-free-scheduling}
\begin{algorithmic}[1]
\State \textbf{Input:} Max\_decode\_tasks: $T$, Num\_nvdec: $N$
\State Initialize \texttt{mutex} $\leftarrow \texttt{init\_mutex}()$
\State Initialize \texttt{condition} $\leftarrow \texttt{init\_condition}()$
\State Initialize \texttt{num\_nvdec\_in\_use} $ \leftarrow 0$
\State Initialize \texttt{threadpool} $\leftarrow \texttt{init\_thread\_pool}(T)$
\For{worker in threadpool in parallel}
    \For{$GOP_s$ in worker}
        \State $mutex.lock()$
        \State // Priority is assigned to $GOP_s$ within the same worker
        \If{$num\_nvdec\_in\_use \leq N$}
            \State $Launch\_async\_decode(GOP_s)$ 
            \State // The remaining $GOP_s$ within the same worker is prioritized
            \State $num\_nvdec\_in\_use \leftarrow num\_nvdec\_in\_use + 1$
        \Else
            \State $condition.wait(mutex)$
        \EndIf
    \EndFor
\EndFor
\end{algorithmic}
\end{algorithm}

\begin{figure}
    \centering
    \includegraphics[width=1\linewidth]{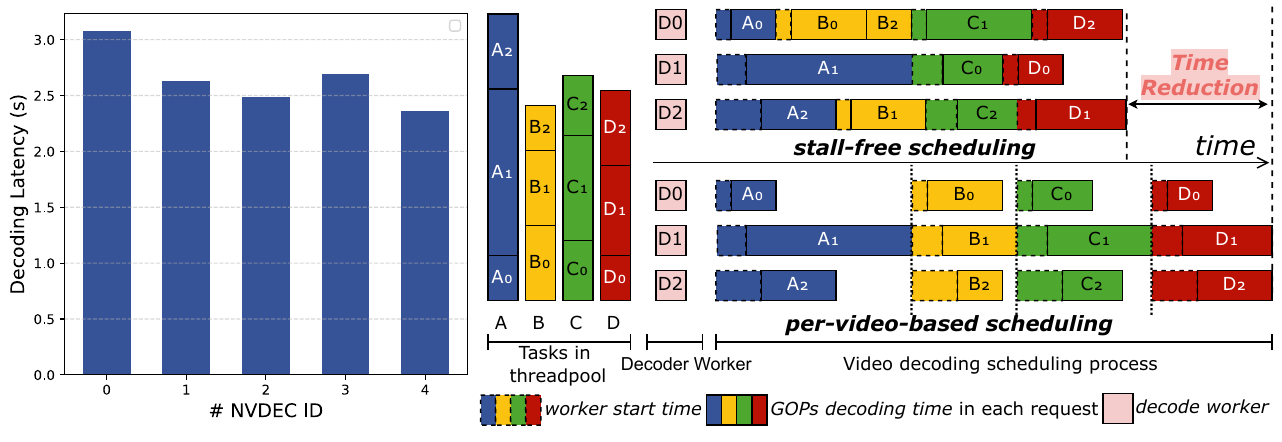}
    \caption{\emph{left}: Decoding times of each NVDEC via profiling on a single A100 GPU. \emph{right}: Comparison between stall-free and default per-video scheduling. }
    \label{fig:stall_free}
     \vspace{-3mm}
\end{figure}

\subsection{Stall-free Scheduling} \label{Stall-Free}
Although GOP-based parallel decoding accelerates video processing by utilizing multiple decoding units, each unit still exhibits periods of underutilization. As shown in Figure \ref{fig:stall_free} (left), NVDEC runtimes fluctuate during decoding; units that finish early must wait for the slowest one, leaving hardware idle and lowering overall efficiency. To mitigate this, we introduce a stall-free scheduling strategy that uses $GOP_s$ as the scheduling granularity rather than treating the entire video as a single unit.
Figure \ref{fig:stall_free} (right)\footnote{We found that worker initialization is serialized—initializing one worker blocks the others—hence the differing initialization times in Figure \ref{fig:stall_free}b.} illustrates the scheduling workflow. Relative to whole-video scheduling (lower part of Figure \ref{fig:stall_free} right), $GOP_s$-level scheduling (upper part) enables immediate dispatch of the next $GOP_s$ segment whenever an NVDEC becomes idle, thereby increasing decoder utilization and improving overall throughput.

Algorithm \ref{alg:stall-free-scheduling} presents the core of our stall-free scheduling. The algorithm selects $T$ workers from the thread pool to run decoding tasks in parallel~(line 6). Each worker attempts to acquire the mutex; once it obtains the lock, it asynchronously schedules its $GOP_s$ segments and releases the lock only after all segments have been dispatched or NVDEC resources are saturated~(lines 7-11, 15). After completing a $GOP_s$, the scheduler prioritizes waking the same worker to dispatch its remaining $GOP_s$ segments~(lines 12-13).

\subsection{Minimizing Frame Pixel Memory Usage. } \label{memory-saving}
Mainstream deep-learning video decoding frameworks (e.g., Decord~\cite{decord_decoder}, TorchCodec~\cite{torchcodec_decoder}) pre-allocate a contiguous GPU memory block upon each decoding request to store decoded frames. This strategy poses little memory pressure when decoding is serial, with a single video as the scheduling unit. In FlashCodec, however, request acceptance and decoding execution are asynchronous, potentially leading to a backlog of decoding requests. To avoid excessive GPU memory pre-allocation and out-of-memory (OOM) errors, FlashCodec allocates GPU memory only after each rank completes decoding. Additionally, the current-worker-prioritized $GOP_s$ scheduling strategy (Algorithm~\ref{alg:stall-free-scheduling}, lines 9 and 12), combined with multi-GPU parallel decoding, further reduces GPU memory usage.

\section{UnifiedServe Design} \label{degisn}
\begin{figure}
    \centering
    \includegraphics[width=1\linewidth]{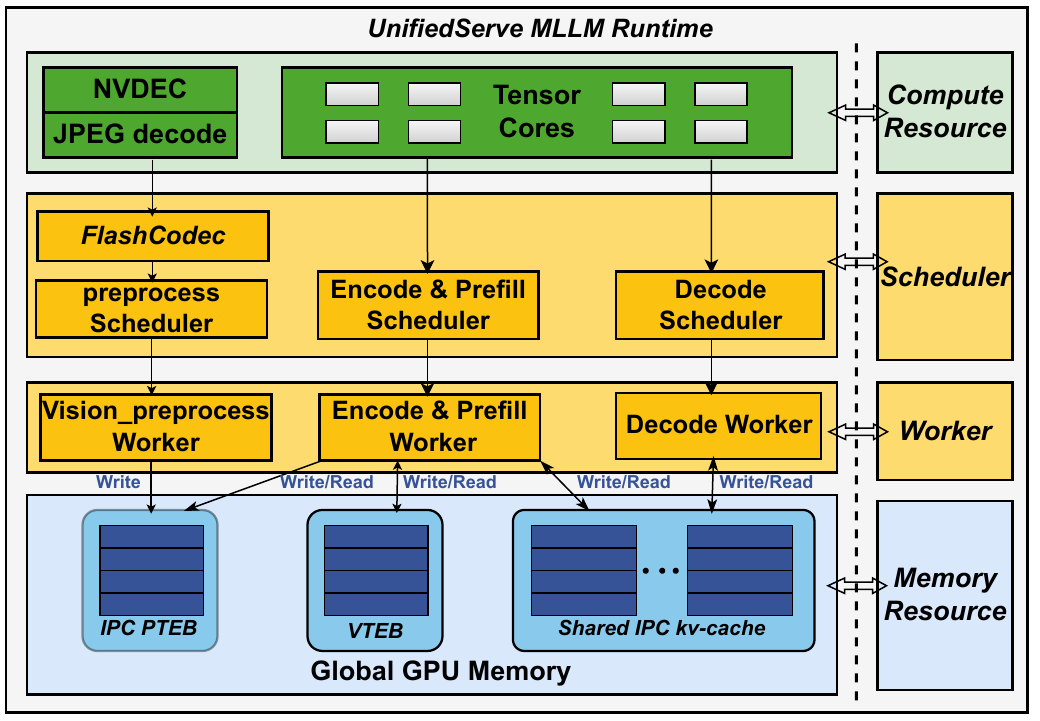}
    \vspace{-2mm}
    \caption{UnifiedServe Overview.}
    \label{fig:overview}
    \vspace{-4mm}
\end{figure}

In this section, we first show the high-level overview of UnifiedServe (\S \ref{sec:overview}). 
We then explain our embedding buffer management (\S\ref{sec::embedding_buffer}), describe how we process the visual modality (\S\ref{sec::vsual_modality_processing}) and orchestrate MLLM generation (\S\ref{sec::mllm_gen}), and detail how these components collaborate (\S\ref{sec::Prefill-encode_Orchestration}).

\subsection{Design Overview}\label{sec:overview}
The architecture of UnifiedServe is shown in Figure~\ref{fig:overview}. 
The key design principle behind UnifiedServe is to allow multiple stages within the MLLM inference to execute in parallel via resource sharing (MPS~\cite{nvidia_mps}), thereby maximizing system throughput while minimizing the blocking of inference caused by the encoding process, and fully utilize system decoding resources to accelerate visual modality decoding.
As shown in Figure \ref{fig:overview}, there are three major workers in the UnifiedServe runtime. 
The first vision\_process worker is responsible for decoding both video and image inputs. It employs FlashCodec (\S~\ref{sec::flashcodec_techniques}) as the visual modality decoder, which leverages all available system decoding resources (e.g., CPUs and GPUs) to accelerate visual modality decoding in parallel (\S~\ref{sec::vsual_modality_processing}).
Additionally, to prevent the encoder from blocking LLM decoding, we abstract the encoder as a dedicated worker and co-locate it with the prefill worker in the same process for joint scheduling (\S~\ref{sec::Prefill-encode_Orchestration}). We discuss the rationale for this co-location in \S~\ref{sec::mllm_gen}. In contrast, the decode worker runs in a separate process. All workers execute in parallel via shared system resources.
The vision\_process worker and the encode worker, as well as the prefill worker and the decode worker, share the same physical memory region to enable efficient inter-process communication (IPC), implemented via the IPC patch buffer and IPC KV-Cache shown in Figure~\ref{fig:overview}.
        
\subsection{Embedding Buffer Management}\label{sec::embedding_buffer}
In UnifiedServe, the interactions between the vision\_process worker and the encode worker, as well as between the encode worker and the prefill worker, are fully asynchronous. In other words, both pairs follow a producer-consumer model in which the producer’s outputs are not immediately consumed. This necessitates an efficient mechanism for storing and managing intermediate results.
Unlike vLLM~\cite{kwon2023efficient} and SGlang~\cite{zheng2024sglang}, which store each request’s intermediate artifacts in a single contiguous memory region, our key design principle is to virtualize the storage space for these intermediate results and materialize them into contiguous memory only at use time. Since both encode and prefill execute only one chunk per iteration (\S\ref{sec::vsual_modality_processing} and \ref{sec::mllm_gen}), this approach yields significantly higher memory efficiency.
UnifiedServe applies classic virtual-memory paging techniques to manage a dedicated buffer for patch/visual-token embeddings, analogous to the paging mechanism used by PagedAttention~\cite{kwon2023efficient} for KV-cache management.

\begin{figure}
    \centering
    \includegraphics[width=1\linewidth]{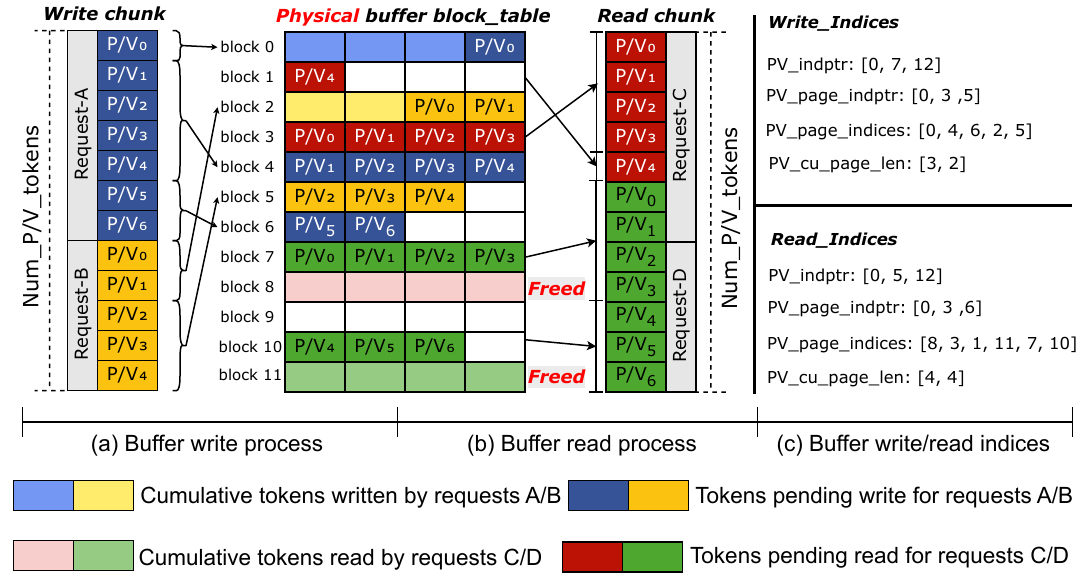}
    \caption{Embedding Buffer Management.
    After each read\_chunk completes in the encode or prefill stage, processed blocks (e.g., 8 and 11) are promptly freed.}
    \label{fig::buffer_management}
\end{figure}

We illustrate the detailed buffer read/write workflow in Figure~\ref{fig::buffer_management}. Similar to FlashInfer’s RaggedIndex~\cite{ye2025flashinfer}, each buffer write/read request in a UnifiedServe iteration carries four indices: \texttt{pv\_indptr}, \texttt{pv\_page\_indptr}, \texttt{pv\_page\_indices} and \texttt{pv\_cu\_page\_len}(Figure~\ref{fig::buffer_management}c).
\texttt{pv\_indptr} stores, for each request, the starting and ending token positions as well as the total number of tokens to be written or read.
\texttt{pv\_page\_indices} records the page IDs used by each request in the current iteration.
\texttt{pv\_page\_indptr} indexes, for each request, the corresponding segment within \texttt{pv\_page\_indices}.
\texttt{pv\_cu\_page\_len} tracks the cumulative number of tokens written or read by each request in all previous iterations.
As shown in Figure~\ref{fig::buffer_management}a and b, for each read or write request, UnifiedServe first determines the number of tokens and their chunk-local positions using \texttt{pv\_indptr}. It then retrieves the page IDs via \texttt{pv\_page\_indices} and \texttt{pv\_page\_indptr}. Next, it uses \texttt{pv\_cu\_page\_len} to obtain the cumulative token count from prior iterations, which in turn determines the starting offset $n$ for the current iteration. Finally, the system reads or writes the required tokens beginning at the $n-th$ token of the pages associated with the request.

After each iteration, both the encode worker and the prefill worker immediately release all pages that have been fully consumed in that iteration (e.g., pages 8 and 11 in Figure~\ref{fig::buffer_management}).

\subsection{Visual Modality Processing}\label{sec::vsual_modality_processing}

\begin{figure*}[t]
    \centering
    \begin{minipage}{0.48\linewidth}
        \centering
        \includegraphics[width=\linewidth]{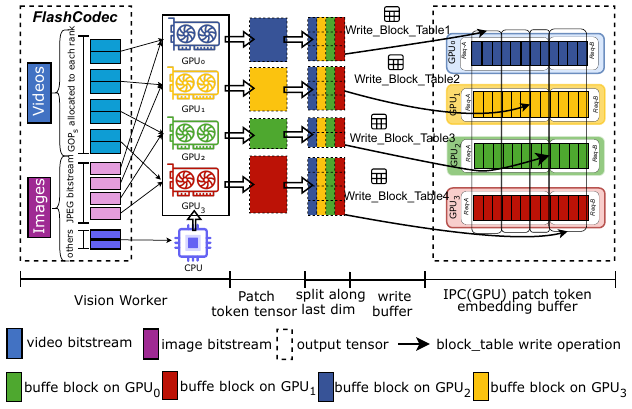}
        \caption{Visual modality decoding process.}
        \label{fig:vision-decoding}
    \end{minipage}
    \hfill
    \begin{minipage}{0.48\linewidth}
        \centering
        \includegraphics[width=\linewidth]{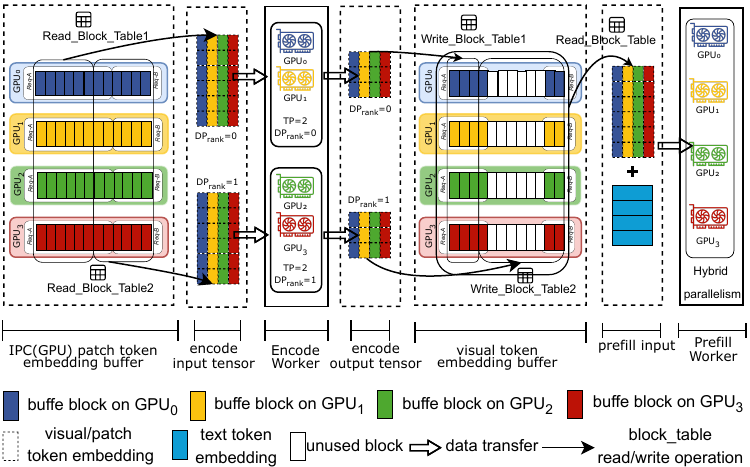}
        \caption{MLLM generation process.}
        \label{LM_generation_process}
    \end{minipage}
\end{figure*}

In UnifiedServe, the vision\_process worker is dedicated to handling the visual modality. Videos and JPEG-encoded images are processed on GPUs, while images in other formats (e.g., PNG) are processed on CPUs. The scheduling of visual modality processing is managed internally by FlashCodec; the vision\_process worker is solely responsible for issuing requests, receiving results, and writing those results into an IPC patch buffer. Video decoding requests employ the stall-free scheduling mechanism described in \S\ref{Stall-Free}, whereas image decoding requests follow a FCFS policy and support batching.
As shown in Figure~\ref{fig:vision-decoding}, to improve GPU memory utilization, the vision\_process worker maintains an IPC patch buffer on each GPU rank. For every computation request assigned to a GPU rank, the resulting patch token embeddings are first generated, then split along the last dimension, and finally each resulting chunk is written into the IPC patch buffer of the corresponding GPU. Because we do not split along the first dimension (i.e., the patch token dimension), a single page\_block\_table suffices to manage the buffers across all GPU ranks.

The workflow of the vision\_process worker follows a multi-producer, single-consumer pattern: each GPU rank produces data that must be written sequentially into the IPC patch buffer. We design a \emph{CollectiveWriteQueue}, which accepts write requests from all ranks, enqueues them, and processes them one at a time. For each request, the queue broadcasts it to the buffer-write subthreads on all ranks; each subthread participates in a collective scatter to obtain its chunk and writes it into its local IPC patch buffer.

\subsection{MLLM Generation}\label{sec::mllm_gen}

\noindent\textbf{LM Decoding.}
To prevent encoding and prefill from blocking decode generation, we deploy the decode worker in a separate process that executes concurrently with other workers through resources sharing.
As shown in Figure~\ref{fig:overview}, the decode worker and the prefill worker share a common IPC-based KV-Cache. The prefill worker only needs to pass scheduling metadata (e.g., block\_table, first\_token\_id) for completed prefill requests to the decode worker, which then processes all pending decode requests in a batched manner.

\noindent\textbf{Patch-token Encoding and LM Prefill.}
There are two strategies for scheduling the encode and prefill workers: (1) running them in two separate processes to enable parallel execution, or (2) colocating them within a single process and interleaving their execution via a scheduling policy.
While the first approach enables concurrency between encode and prefill, both stages are computationally intensive. Moreover, some MLLMs achieve superior accuracy by employing substantially larger vision encoders (e.g., a 5B ViT in InternVL3~\cite{zhu2025internvl3} and a 22B ViT in PaLI-X~\cite{beyer2024paligemma} and PaLM-E~\cite{beyer2024paligemma}). In such cases, concurrent execution of encode and prefill leads to severe contention for GPU compute and memory bandwidth, dramatically increasing decode-generation latency. To mitigate this interference, we adopt the second strategy—co-scheduling both workers within a single process—thereby ensuring predictable resource usage and minimizing tail latency during generation. We discuss the scheduling policy for the encode and prefill workers in detail in \S\ref{sec::Prefill-encode_Orchestration}.

As shown in Figure~\ref{LM_generation_process}, UnifiedServe supports hybrid-parallel inference (combining data and tensor parallelism) for the vision encoder. 
The encode worker first reads its input from the IPC patch buffer according to a provided \texttt{read\_block\_table} and supports chunked encoding. 
Since ViTs typically operate on a per-image basis rather than per-token:
all tokens originating from the same image attend to each other via full self-attention. Consequently, UnifiedServe chunks inputs at the image granularity rather than the token level. Finally, the encode worker writes its output into the visual token buffer using a scheduler-assigned \texttt{write\_block\_table}, making the results available to the prefill stage.
The prefill worker then reads the visual token embeddings for the currently scheduled batch from the visual token buffer using the \texttt{read\_block\_table} and concatenates them with text token embeddings to form the joint multimodal input for the prefill. 

\subsection{Prefill-encode Orchestration}\label{sec::Prefill-encode_Orchestration}
To avoid severe resource contention caused by concurrent execution of two compute-intensive workers and to enable finer-grained control over GPU resource utilization by compute-intensive tasks, UnifiedServe schedules encoding and prefill within the same process (\S\ref{sec::mllm_gen}).
The prefill-encode scheduling orchestration is presented in Algorithm~\ref{Prefill-encode-Orchestration}, where the execution of encode and prefill blocks each other.

UnifiedServe first gets the budget of maximum number of tokens that can be executed in an encode and prefill batch based on user specified SLO(lines 1-3). We set the token budget empirically based on the considerations in Sarathi~\cite{agrawal2024taming}.
In every scheduling iteration, we first incorporate any partially completed prefill requests (lines 8-11). Only after all currently running requests have been accommodated do we admit new requests (lines 12-28). Prior to admitting a new request, we first determine whether it is a multimodal request and whether its encoding has completed (line 13). If the request is multimodal and its encoding has not yet finished, we add it—along with other pending encoding requests—to the batch, ensuring that the total number of encoding tokens in the batch is no less than the encoding token budget (lines 14-22). When adding prefill requests to the batch, we compute the maximum chunk size that can fit within the remaining token budget for that batch (lines 10,24).
The number of tokens in an encoding batch may exceed the encoding token budget; consequently, encoding may need to be executed multiple times. Moreover, if a prefill operation is blocked due to insufficient KV cache allocation, encoding can still proceed (line 29). In contrast, the number of tokens in a prefill batch is guaranteed not to exceed the prefill token budget, and thus requires only a single execution (line 30).

By serializing the execution of prefill and encoding and restricting their computational load in each scheduling iteration, the system ensures that the generation latency of decode steps can be effectively controlled through adjusting the token budgets allocated to prefill and encoding.

\begin{algorithm}[t]
\footnotesize 
\caption{\textbf{Prefill-encode Orchestration}.}
\label{Prefill-encode-Orchestration}
\begin{algorithmic}[1]
\State \textbf{Input:} $T_{\max}$, Application TBT SLO.
\State Initialize \texttt{p\_token\_budget:} $\tau \leftarrow \texttt{get\_p\_token\_budget}(T_{\max})$
\State Initialize \texttt{e\_token\_budget:} $\alpha \leftarrow \texttt{get\_e\_token\_budget}(T_{\max})$
\State Initialize \texttt{batch\_num\_tokens:} $n_p \leftarrow 0$, $n_e \leftarrow 0$
\State Initialize \texttt{current batch:} $B \leftarrow \emptyset$
\State Initialize \texttt{encode-finished batch:} $E \leftarrow \emptyset$
\While{True}
    \For{$R$ in $B$}
        \If{not $is\_prefill\_complete(R)$}
            \State $c \leftarrow get\_next\_chunk\_size(R, \tau, n_p)$
            \State $n_p \leftarrow n_p + c$
        \EndIf
    \EndFor
    \State $R_{new} \leftarrow get\_next\_request()$
    \If{$is\_multimodal\_req(R_{new}) \land R_{new} \notin E$}
        \State $R_{ref} \leftarrow R_{new}$
        \While{True}
            \If{$finished\_visual\_modality(R_{ref})$}
            \If{$allocated\_or\_can\_alloc\_buffer(R_{ref}) \land n_e < \alpha$}
                \State $p \leftarrow get\_patch\_token\_size(R_{ref})$
                \State $n_e \leftarrow n_e + p$, $E \leftarrow R_{ref}$
            \Else
                \State \textbf{break}
            \EndIf
            \EndIf
            \State $R_{ref} \leftarrow get\_next\_multimodal\_request\_reference()$
        \EndWhile
    \EndIf
    \While{$allocated\_or\_can\_alloc\_cache(R_{new}) \land n_p < \tau$}
        
        \State $c \leftarrow get\_next\_chunk\_size(R_{new}, \tau, n_p)$
        \If{$c > 0$}
            \State $n_p \leftarrow n_p + c$, $B \leftarrow R_{new}$
        \Else
            \State \textbf{break}
        \EndIf
    \EndWhile
    \State $chunked\_encode\_hybrid\_batch(E)$
    \State $process\_prefill\_hybrid\_batch(B)$
    \State $B, E \leftarrow filter\_finished\_requests(B, E)$
    \State $n_p \leftarrow 0$, $n_e \leftarrow 0$
\EndWhile
\end{algorithmic}
\end{algorithm}

\section{Implementation}
FlashCodec extends TorchCodec~\cite{torchcodec_decoder} with 5.6K lines of C++/CUDA code and provides interfaces for image decoding and frame resizing via selectable interpolation algorithms. JPEG is decoded via dedicated hardware, while other formats are decoded on the CPU.
FlashCodec offers three primary APIs: 
(\textit{i})~\texttt{analyse\_bitstream} assigns a unique \texttt{key\_id} to each request and returns it along with video metadata and the number of GPUs allocated. 
(\textit{ii})~\texttt{add\_decoding\_request} enqueues a decoding request for processing in a thread pool. 
(\textit{iii})~\texttt{get\_decoding\_output} returns the completed \texttt{key\_id} and its corresponding GPU tensor, blocking the caller until the result is ready while releasing the GIL during the wait.

We build UnifiedServe on Sarathi-Serve~\cite{agrawal2024taming} and extend it to support MLLM inference (e.g., Qwen2.5-VL-Serious~\cite{Qwen2.5-VL}, Internvl3-Serious~\cite{zhu2025internvl3}). The vision worker employs FlashCodec for video and image decoding. We implement custom kernels to manage IPC patch/vision buffers and enable sharing of buffers, model parameters, and the KV cache across processes. To minimize inter-worker communication and synchronization overhead, we introduce shared-memory-based message passing and synchronization primitives, and use NCCL~\cite{nccl} for both IPC buffer transfers and parallel inference communication.

\section{Evaluation}
\label{evaluation}

\begin{table}[t]
\centering
\scriptsize
\renewcommand{\arraystretch}{1.05}
\setlength{\tabcolsep}{2.5pt}

\begin{tabularx}{0.48\textwidth}{l|c|c|c|>{\centering\arraybackslash}X}
\hline
\textbf{Models} &
\textbf{\makecell{GPU\\Config}} &
\textbf{\makecell{Vision\\Encoder}} &
\textbf{\makecell{LM\\Decoder}} &
\textbf{\makecell{Memory\\Capacity}} \\
\hline
Qwen2.5-VL-32B~\cite{Qwen2.5-VL}  & 4 $\times$ A100s  & 0.5B & 32B & 80 $\times$ 4 GB \\
InternVL3-38B~\cite{zhu2025internvl3}   & 4 $\times$ A100s  & 6B & 32B & 80 $\times$ 4 GB \\
\hline
\end{tabularx}

\caption{Models and GPU configurations.}
\label{exp_cfg}
\end{table}

\subsection{Environment Setup} \label{e0}
\textbf{Testbed.} 
We deploy UnifiedServe in the experimental environment summarized in Table~\ref{exp_cfg}. 
The monolithic system uses TP=4, while the split-based system applies TP=2 both the encode and prefill, and TP=2 for the decode. In UnifiedServe, TP=4 is used for the LM backend, DP=2 + TP=2 for Qwen2.5-VL encoding, and a TP=4 configuration for InternVL3 encoding.
The baseline uses a token budget of 2048; UnifiedServe uses 2048 for prefill and 10240 for Qwen2.5-VL encode and 5120 for InternVL3 encode. Both videos and images are resized to a resolution of 224 $\times$ 224.

\begin{table}[t]
\centering
\scriptsize
\renewcommand{\arraystretch}{1.05}
\setlength{\tabcolsep}{2.5pt}

\begin{tabularx}{0.48\textwidth}{l|c|>{\centering\arraybackslash}X|c}
\hline
\textbf{Dataset} & 
\textbf{Type} & 
\textbf{Description} & 
\textbf{Duration} \\
\hline
MLVU~\cite{zhou2024mlvu} & Video & Multi-task annotated long videos. & 8-10min \\
EgoSchema~\cite{mangalam2023egoschema} & Video & Multi-task annotated short videos. & 3min \\
VisionArena~\cite{chou2025visionarena} & Image & Images with paired text descriptions. & - \\
\hline
\end{tabularx}

\caption{Datasets for workload generation.We excluded videos from MLVU with excessively long durations and retained only those with durations between 8 and 10 minutes.}
\label{exp_ds}
\end{table}

\textbf{Workloads.} We select representative real-world multimodal datasets, including videos and images of various sizes. The benchmarked datasets are MLVU~\cite{zhou2024mlvu}, EgoSchema~\cite{mangalam2023egoschema}, and VisionArena~\cite{chou2025visionarena}, as shown in Table~\ref{exp_ds}. The MLVU dataset provides long video sequences for multi-task video understanding, EgoSchema contains short video clips, and VisionArena pairs images with textual descriptions for vision-language tasks.
We use FlashCodec only when evaluating on the MLVU dataset, and adopt the baseline's decord decoder for EgoSchema to assess UnifiedServe’s performance without FlashCodec.

\noindent\textbf{Approaches for Inference Framework.}
For comparison with \emph{UnifiedServe}, we evaluated the following approaches: 

\begin{itemize}
    [left=0pt]
    \item \textbf{vLLM~\cite{kwon2023efficient}.} A unified system accompanied by multiple open-source optimizations. We compare with its chunked-prefill~\cite{agrawal2024taming, Deepspeed_mii} based implementation (vLLM-s) and pd-disaggregation~\cite{zhong2024distserve, patel2024splitwise} based implementation (vLLM-d).For a fair comparison, we replace vLLM’s cv2-based video decoder with the more efficient decord to match the setup used by sglang.

    \item \textbf{SGLang~\cite{zheng2024sglang}.}~Another unified system with multiple open-source optimizations.~We also compare with its chunked-prefill based implementation (SGLang-s) and PD-disaggregation based implementation (SGLang-d).
\end{itemize}

\noindent\textbf{Approaches for video decoder.}
For comparison with \emph{FlashCodec}~\footnote{Starting from v0.8.0, torchcodec~\cite{torchcodec_decoder} replaces the FFmpeg~\cite{ffmpeg} backend with the Beta backend for video decoding, achieving higher speed. flashcodec supports both FFmpeg and Beta backends. To ensure comparability with the baseline, we employ the FFmpeg backend in all experiments.}, we evaluated the following approaches: 

\begin{itemize}
    [left=0pt]
    \item \textbf{Decord~\cite{decord_decoder}.}  A multimedia loading framework designed for machine learning applications. Decord remains integrated into popular libraries like Hugging Face’s Transformers~\cite{wolf-etal-2020-transformers} and, by extension, inference frameworks such as vLLM~\cite{kwon2023efficient} and SGLang~\cite{zheng2024sglang}, providing both CPU and CUDA-based video decoding.

    \item \textbf{TorchCodec~\cite{torchcodec_decoder}.} A work-in-progress library from the PyTorch team designed to offer faster multimedia processing than TorchVision~\cite{torchvision2016}, with support for both CPU and CUDA-accelerated video decoding via TorchCodec.

    \item \textbf{DeepCodec~\cite{schneider2025quickvideo}.} A CPU-only video decoder that boosts performance by reducing threads per frame and decoding more GOPs in parallel—rather than accelerating individual frames—yielding significant speedup on many-core CPUs.
\end{itemize}

\noindent\textbf{Metrics.} Our metrics primarily focus on the average, P99, and P95 values for TTFT and TBT latency. Additionally, we evaluate the per-request E2E latency, request throughput under specific SLO constraints, and assess both SLO attainment and scalability.

\begin{figure*}
    \centering
    \includegraphics[width=1\linewidth]{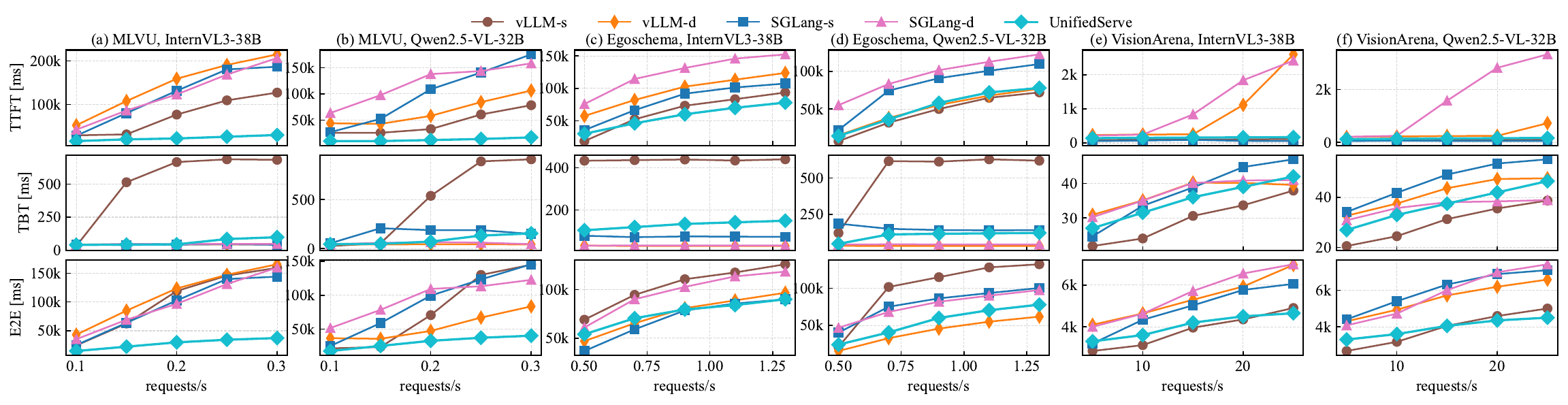}
    \caption{Overall comparison of the average TTFT, TBT, and E2E latency of UnifiedServe, vLLM, and SGLang across two models and three datasets.}
    \label{exp1}
\end{figure*}

\subsection{Overall Performance}\label{e1}

We evaluated the overall performance of UnifiedServe, vLLM, and SgLang across two models and three types of multimodal datasets. 
Figure~\ref{exp1} compares the average TTFT, TBT latency, and E2E time per request at different request sending rates. For long videos (MLVU), the prolonged video decoding and encoder computation pose significant bottlenecks for existing baselines. 

In vLLM with chunked prefill, although its average TTFT is the lowest, only 50\% of the Split-Based method, its average TBT latency can exceed 300ms due to severe blocking of the decode chunks by encoder computations. 
In contrast, Split-Based methods, such as vLLM-D and sglang-d, achieve outstanding TBT latency (averaging below 20ms) because decoding is handled by completely independent GPUs, avoiding any blocking. However, the heavy video decoding, prefill, and encode processes result in relatively poor TTFT latency for these methods. 
The scheduling characteristics of sglang-s cause it to adopt a chunked prefill execution mode under low request rate pressure (resulting in low TTFT and high TBT). However, under high request rate pressure, it falls back to a serial execution mode (leading to high TTFT and low TBT). 
In contrast, UnifiedServe achieves high performance in both TTFT and TBT. Compared to vLLM-D, our average TTFT is reduced by 80\%, while the TBT latency only increases by 50\% approximately. This is because UnifiedServe leverages resource sharing to fully utilize all compute resources while avoiding blocking. 
Additionally, our FlashCodec enables lower TTFT without significantly impacting TBT, unlike vLLM-S.

In the short video scenario, UnifiedServe uses the same decord video decoder as the baseline to show performance without FlashCodec.
The overall performance of the baselines is similar to that observed for long videos. Because video decoding time and TTFT are both shorter than in long-video scenarios, monolithic-based systems are more prone to generation stalls at the same request rate. Compared with monolithic-based systems, UnifiedServe achieves a very similar TTFT but a smaller TBF latency, resulting in a shorter E2E latency. Compared with split-based systems, although its TBT latency remains higher, UnifiedServe attains a smaller TTFT because the first-token generation can leverage all system resources, and multimodal preprocessing and encoding increase the first-token time, thereby reducing the impact of the decode stage on other stages at the same request rate.

In the image scenario, due to the reduced pressure on the encoder, monolithic-based systems no longer experience significant blocking on TBT. 
In contrast, split-based systems suffer from suboptimal resource utilization during the prefill and decode stages, resulting in worse TTFT and TBT performance compared to the monolithic-based systems. 
UnifiedServe, on the other hand, fully leverages all available compute resources for parallelism while minimizing the blocking caused by the encoder. 
As a result, our overall performance is comparable to vLLM-s, and our TTFT at higher request rates outperforms split-based systems.
Additionally, in the majority of scenarios, UnifiedServe achieves the shortest average E2E time per request. This is because our FlashCodec significantly reduces the latency of video/image decoding, and the resource-sharing scheduling ensures that the prefill and decode processes of each request are processed promptly, thus avoiding any blocking.

\begin{figure}
    \centering
    \includegraphics[width=1\linewidth]{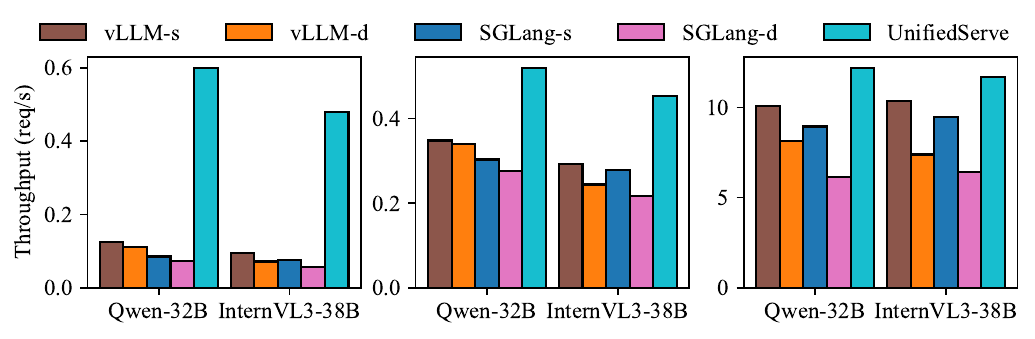}
    \caption{Overall comparison of UnifiedServe's average throughput across different datasets (From left to right: MLVU, EgoSchema, and VisionArena).}
    \label{exp3}
\end{figure}

We evaluated the maximum inference throughput of the service across different scenarios, as shown in Figure~\ref{exp3}. It is evident that UnifiedServe achieves the highest throughput, outperforming the current monolithic-based methods by up tp $4.4\times$ on the MLVU dataset. This is because FlashCodec can leverage the decoding resources of all GPUs in the system in parallel to accelerate video decoding and our comprehensive scheduling enables the encoder to run in parallel with decoding, thus increasing parallelism. In contrast, split-based scheduling methods exhibit lower throughput, achieving only 14\% to 51\% of UnifiedServe’s performance.

\begin{figure}
    \centering
    \includegraphics[width=1\linewidth]{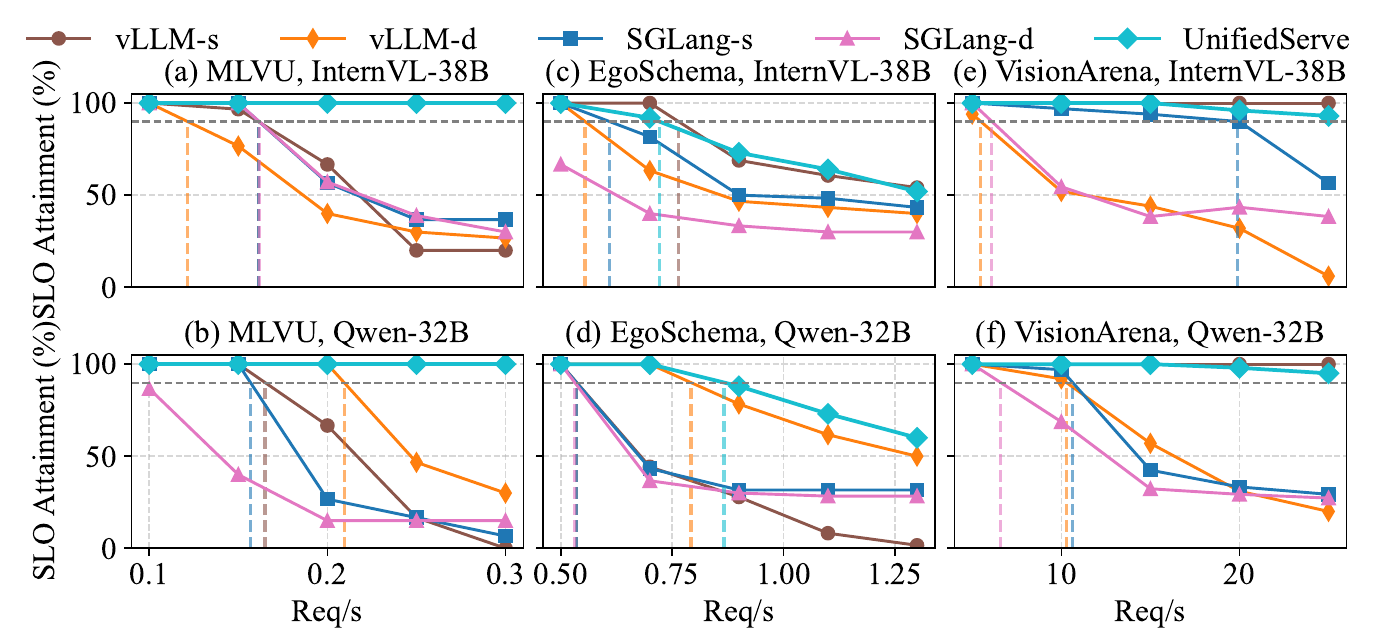}
    \caption{Comparison of SLO Attainment under Different Request Rates.}
    \label{exp2}
\end{figure}

\begin{table}[t]
\centering
\scriptsize
\renewcommand{\arraystretch}{1.1}
\setlength{\tabcolsep}{5pt}

\begin{tabular}{l|cc|cc|cc}
\hline
\multirow{2}{*}{\textbf{Models}} &
\multicolumn{2}{c|}{\textbf{MLVU}} &
\multicolumn{2}{c}{\textbf{Egoschema}} &
\multicolumn{2}{c}{\textbf{VisionArena}}\\
\cline{2-7}
 & \textbf{TTFT} & \textbf{TBT} & \textbf{TTFT} & \textbf{TBT} & \textbf{TTFT} & \textbf{TBT}\\
\hline
Qwen2.5-VL-32B & 80 & 0.7 & 80 & 0.6 & 0.25 & 0.05 \\
InternVL3-38B & 140 & 0.8 & 100 & 0.7 & 0.25 & 0.05 \\
\hline
\end{tabular}

\caption{Latency SLOs under different workloads }
\label{SLOs}
\end{table}

\subsection{SLO Attainment}\label{e2}
In this section, We first evaluate SLO attainment for TTFT and TBT using the settings in Table~\ref{SLOs}, and then evaluate SLO attainment under a fixed RPS across varying SLOs. Our goal is to measure the inference service’s capacity-specifically, the request rate it can sustain while meeting these SLOs.

Figure~\ref{exp2} shows the maximum request capacity under different SLO attainments. In the long video request scenario, UnifiedServe increases the request capacity while still meeting the SLO for 100\% of requests. This improvement arises because both monolithic-based and split-based systems struggle to meet both TTFT and TBT SLOs simultaneously in long-video scenarios, resulting in lower req/s capacity. UnifiedServe, however, maintains TTFT SLO through efficient decoding and high-priority prefill/encode, while improving TBT SLO by parallelizing decode computation through resource sharing and avoiding blocking between decode steps. 
In short-video request scenarios, UnifiedServe achieves the highest SLO attainment in most cases, even without employing FlashCodec to accelerate video decoding.
In image request scenarios, vLLM-S achieves the best SLO attainment, likely because image decoding incurs relatively low overhead and is less prone to becoming a bottleneck. Moreover, the resized resolution is small (At a resolution of 224 $\times$ 224, Qwen-2.5-VL yields 128 patch tokens and 32 visual tokens per image), resulting in fewer visual tokens and thereby substantially reducing the impact of vision encoding on language model generation.

\begin{figure}
    \centering
    \includegraphics[width=1\linewidth]{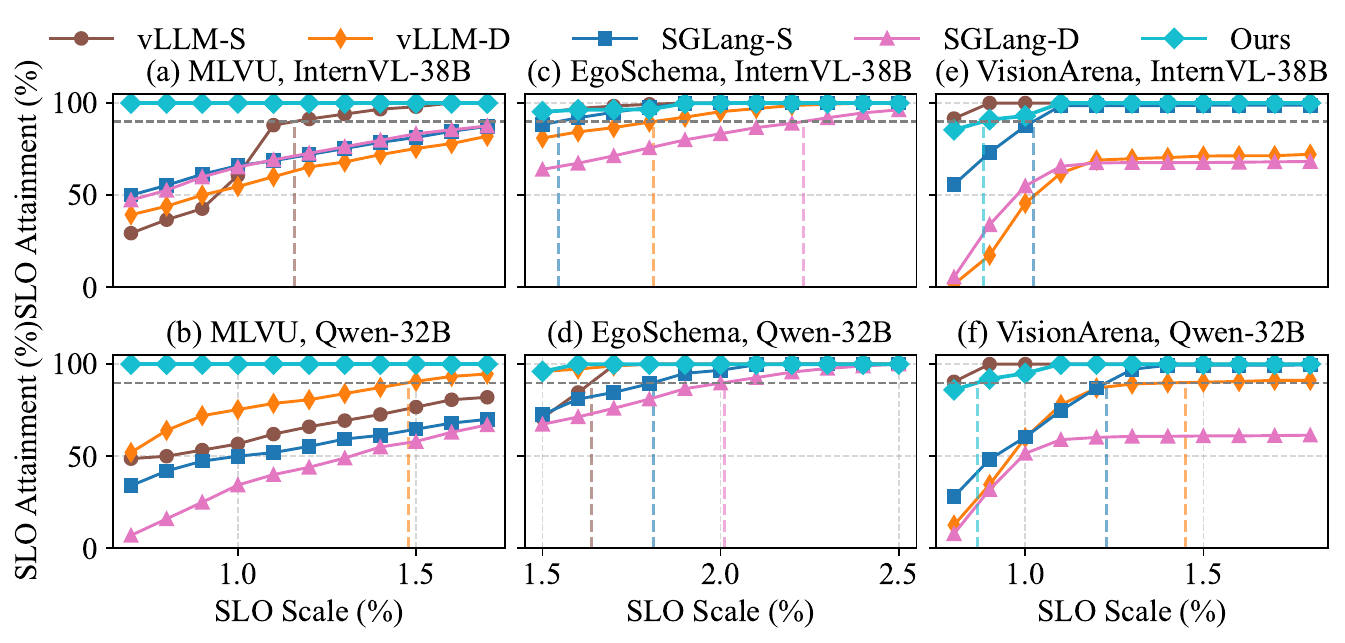}
    \caption{Comparison of SLO attainment  under different SLO scales.}
    \label{exp5}
\end{figure}

Figure~\ref{exp5} shows the SLO attainments under different SLO scales.
Compared to monolithic-based and split-based systems, UnifiedServe achieves several times more stringent SLO requirements. Due to the longer video decoding and encoder lengths in the MLVU dataset, split-based systems struggles to meet the TTFT SLO, while monolithic-based systems fails to meet the TBT SLO. 
Although the TTFT SLO for split-based systems is more relaxed for image datasets, UnifiedServe and monolithic-based systems still achieve approximate 2 times more stringent SLO requirements.

\begin{figure}
    \centering
    \includegraphics[width=1\linewidth]{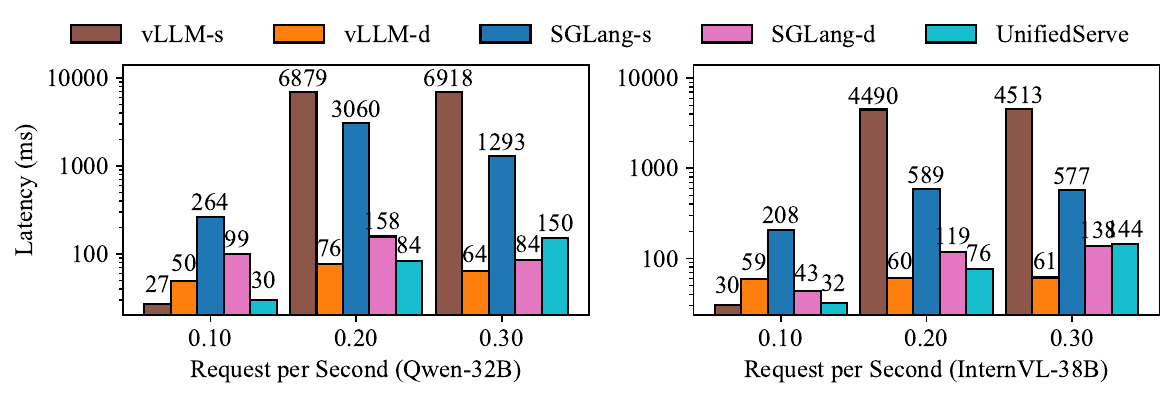}
    \caption{P99 TBT Latency Comparison.}
    \label{exp6}
\end{figure}

Under more stringent SLO attainment target P99, UnifiedServe also demonstrates superior TBT latency, as shown in Figure~\ref{exp6}. Compared to monolithic-based systems, UnifiedServe achieves a reduction in latency by a factor of 83\% for both P99 TBT. In monolithic-based systems, partial blocking caused by the encoder (over 500ms) severely slows down the P99 TBT latency, whereas UnifiedServe avoids such blocking delays, ensuring better performance in TBT. 
split-based systems demonstrate the best P99 TBT latency among all baselines, highlighting the advantage of separated decoding, which allows for extremely low and stable TBT latency without interference from other computations. 

\subsection{Video decoding capacity.}\label{e4}

\begin{figure}
    \centering
    \includegraphics[width=1\linewidth]{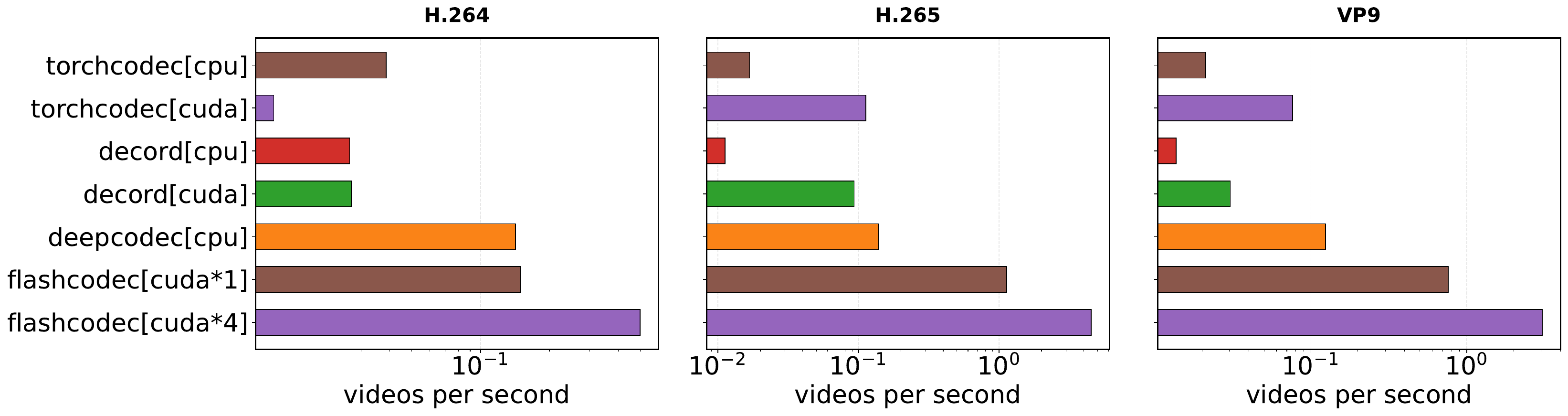}
    \caption{Video Decoding Comparison.The test set comprises 100 videos, with a duration P95 of 20 minutes and a maximum length of 31 minutes. All H.265 and VP9 videos were transcoded from H.264 originals.
    }
    \label{exp_decode}
\end{figure}

We compare FlashCodec with current mainstream video decoders (\S\ref{e0}) in terms of decoding videos with different lengths and encoding formats, as shown in Figure~\ref{exp_decode}. DeepCodec performs decoding exclusively on the CPU, whereas the other decoders are evaluated using both CPU and CUDA-based decoding. FlashCodec significantly outperforms other methods in decoding latency across all video encoding formats and GPU types. Compared to the SOTA method DeepCodec, we achieve up to $9\times$ optimization in decoding latency on 4 A100 GPUs. In terms of decoding latency, even in the H.264 scenario where CPUs have an advantage, DeepCodec still exhibits 4 times higher latency than FlashCodec on 4 $\times$ A100 GPUs. This is because FlashCodec fully leverages all decoding resources on the GPU to decode a single video, resulting in the best latency performance.

\section{Related Work and Discussion.}

\noindent\textbf{Algorithm-oriented MLLM Optimization.}
Existing algorithmic techniques primarily improve multimodal LLM inference efficiency along two axes: \emph{(i)} KV-cache optimization and \emph{(ii)} visual token deduplication.
Methods such as ReKV~\cite{di2025streaming} offload KV caches to CPU memory and fetch only a fixed-size subset most relevant to the current tokens. StreamAgent~\cite{yang2025streamagent} adopts a similar idea but adjusts the retrieval size dynamically. Inf-MLLM~\cite{ning2024inf} reduces KV-cache memory usage via token caching and attention bias, while Elastic Cache~\cite{liu2024efficient} applies importance-based cache merging to conserve GPU memory.
Approaches including Flash-VStream~\cite{zhang2024flash}, Dynamic-LLaVA~\cite{huang2024dynamic} remove redundant visual tokens on the fly to reduce compute and memory costs. These techniques introduce slight accuracy trade-offs but remain orthogonal to system-level optimizations.

\noindent\textbf{MLLM Serving Optimization.} 
To reduce interference during MLLM inference, concurrent works adopt a broadly similar phase-decoupling paradigm to meet SLOs. EPD~\cite{singh2024efficiently} and HydraInfer~\cite{dong2025hydrainfer} deploy encoders on separate instances and support multiple partitioning strategies; ModServe~\cite{qiu2025modserve} further introduces stage-aware model configuration; RedServe~\cite{guo2025rserve} improves parallelism across decoupled stages via intra- and inter-request pipelining.
In contrast, our method decouples phases logically rather than physically, allowing all phases to share system resources. This design delivers higher aggregate throughput while still meet SLOs. Additionally, we optimize multimodal preprocessing—an aspect neglected in prior systems—ensuring SLO compliance even under heavy multimodal input loads.

\textbf{LLM Serving Optimization.} Recently, there has been a growing body of work on efficient LLM serving. Sarathi~\cite{agrawal2024taming} and Orca~\cite{yu2022orca} focus on scheduling and batching in chunked prefill and continuous batching. DistServe\cite{zhong2024distserve}, Splitwise\cite{patel2024splitwise}, TetriInfer\cite{hu2024inference} disaggregate prefill and decoding across heterogeneous resources, aiming to satisfy tail-latency SLOs. vLLM~\cite{kwon2023efficient}, DéjàVu\cite{strati2024d}, Mooncake\cite{qin2024mooncake} and FastServe~\cite{wu2023fast} optimize end-to-end serving with KV-cache scheduling. 
Similar to UnifiedServe, Semi-PD~\cite{hong2025semi} and Nexus~\cite{shi2025nexus} disaggregate prefill and decode and exploit phase-wise resource sharing to boost throughput, whereas our work goes beyond text-only LLMs and targets the more complex MLLM serving scenario.

\section{Conclusion}
FlashCodec and UnifiedServe jointly optimize the end-to-end MLLM pipeline. FlashCodec delivers low-latency, high-throughput multimodal preprocessing via coordinated multi-GPU video decoding. UnifiedServe eliminates cross-stage blocking while maximizing GPU utilization through shared-resource execution. Together, they enable MLLM workloads to meet strict TTFT/TBT SLOs and achieve scalable, high-throughput serving. our proposed framework forms an end-to-end optimized stack that can serve up to 3.0$\times$ more requests or enforce 1.5$\times$ tighter SLOs, while achieving up to 4.4$\times$ higher throughput compared to state-of-the-art systems.

\newpage
{\footnotesize \bibliographystyle{acm}
\bibliography{sample}}


\end{document}